\documentclass[manuscript,screen]{acmart}
\PassOptionsToPackage{table,xcdraw}{xcolor}

\AtBeginDocument{%
  \providecommand\BibTeX{{%
    \normalfont B\kern-0.5em{\scshape i\kern-0.25em b}\kern-0.8em\TeX}}}

\setcopyright{none}
\acmYear{2018}
\acmDOI{XXXXXXX.XXXXXXX}

\acmConference[Conference acronym 'XX]{Make sure to enter the correct
  conference title from your rights confirmation emai}{June 03--05,
  2018}{Woodstock, NY}
\usepackage{booktabs}
\usepackage{graphicx}
\usepackage[table,xcdraw]{xcolor}

\usepackage{multicol}
\usepackage{amsmath}
\usepackage{hyperref}
\usepackage{dsfont}
\usepackage{algorithm}
\usepackage{algpseudocode}
\usepackage{newtxmath}

\usepackage{multirow} 
\usepackage{array} 
\usepackage{colortbl}
\usepackage{caption}
\usepackage{subcaption}

\usepackage[utf8]{inputenc}
\usepackage{enumitem}
\newcommand{\squishlist}{\begin{itemize}[itemsep=1pt,parsep=2pt,topsep=3pt,partopsep=0pt,leftmargin=0em, itemindent=1em,labelwidth=1em,labelsep=0.5em]}
\newcommand{\squishend}{\end{itemize}}

\begin{document}

\title[DeltaLCA]{DeltaLCA: Comparative Life-Cycle Assessment for Electronics Design} 

\author{\href{mailto:zzhihan@cs.washington.edu}{Zhihan Zhang}}
\authornote{These authors contributed equally to this research.}
\email{zzhihan@cs.washington.edu}
\author{Felix H{\"a}hnlein}
\authornotemark[1]
\author{Yuxuan Mei}
\authornotemark[1]
\affiliation{%
  \institution{Paul G. Allen School of Computer Science \& Engineering, University of Washington}
  \city{Seattle}
  \state{WA}
  \country{USA}
}

\author{Zachary Englhardt}
\affiliation{%
  \institution{Paul G. Allen School of Computer Science \& Engineering, University of Washington}
  \city{Seattle}
  \state{WA}
  \country{USA}}

\author{Shwetak Patel}
\affiliation{%
  \institution{Paul G. Allen School of Computer Science \& Engineering, University of Washington}
  \city{Seattle}
  \state{WA}
  \country{USA}}

\author{\href{mailto:adriana@cs.washington.edu}{Adriana Schulz}}
\affiliation{%
  \institution{Paul G. Allen School of Computer Science \& Engineering, University of Washington}
  \city{Seattle}
  \state{WA}
  \country{USA}}

\author{\href{mailto:vsiyer@uw.edu}{Vikram Iyer}}
\affiliation{%
  \institution{Paul G. Allen School of Computer Science \& Engineering, University of Washington}
  \city{Seattle}
  \state{WA}
  \country{USA}}

\definecolor{BrewerGreen}{HTML}{1b9e77}
\definecolor{BrewerOrange}{HTML}{d95f02}
\renewcommand{\shortauthors}{Zhang et al.}
\newcommand{\todo}[1]{\textcolor{blue}{(TODO: #1)}}
\newcommand{\FH}[1]{\textcolor{BrewerGreen}{(FH: #1)}}
\newcommand{\ZZ}[1]{\textcolor{BrewerOrange}{(ZZ: #1)}}

\begin{abstract}
Reducing the environmental footprint of electronics and computing devices requires new tools that empower designers to make informed decisions about sustainability during the design process itself. This is not possible with current tools for life cycle assessment (LCA) which require substantial domain expertise and time to evaluate the numerous chips and other components that make up a device. We observe first that informed decision-making does not require absolute metrics and can instead be done by comparing designs. Second, we can use domain-specific heuristics to perform these comparisons. We combine these insights to develop DeltaLCA, an open-source interactive design tool that addresses the dual challenges of automating life cycle inventory generation and data availability by performing comparative analyses of electronics designs. Users can upload standard design files from Electronic Design Automation (EDA) software and the tool will guide them through determining which one has greater carbon footprint. DeltaLCA leverages electronics-specific LCA datasets and heuristics and tries to automatically rank the two designs, prompting users to provide additional information only when necessary. We show through case studies DeltaLCA achieves the same result as evaluating full LCAs, and that it accelerates LCA comparisons from eight expert-hours to a single click for devices with \textasciitilde30 components, and 15 minutes for more complex devices with \textasciitilde100 components.
\end{abstract}




\begin{teaserfigure}
  \centering
  \includegraphics[width=\linewidth]{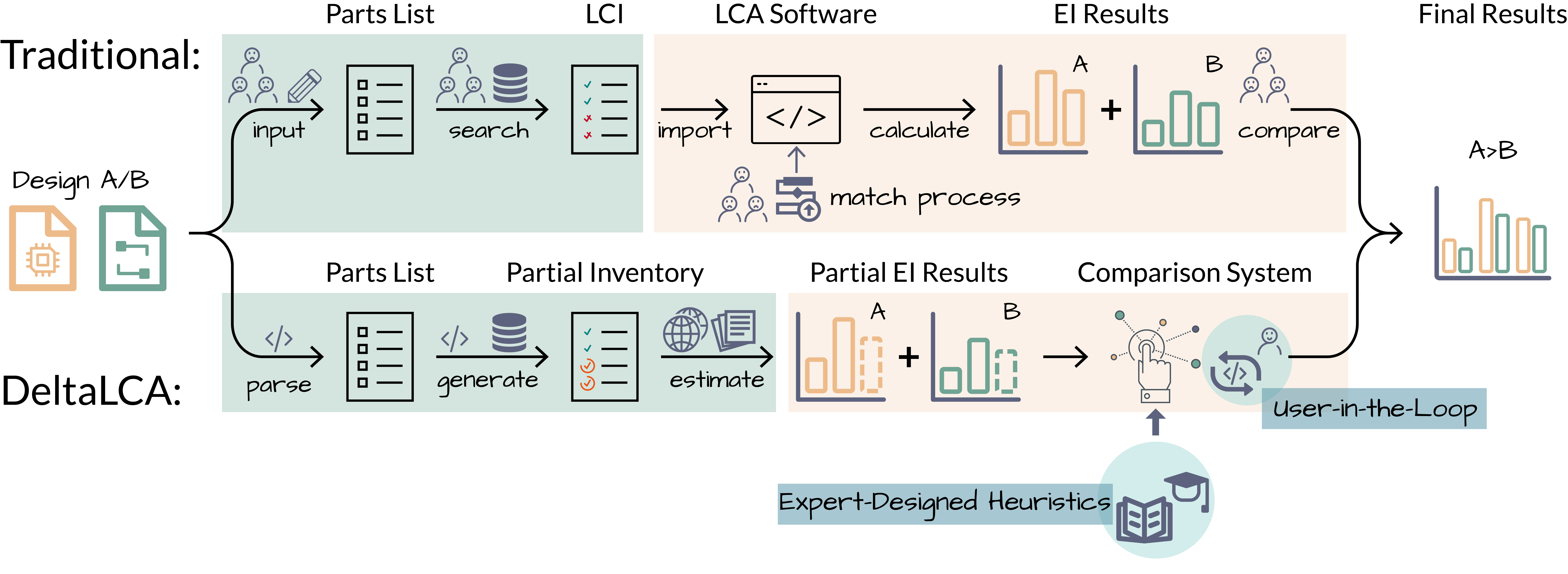}
  \vskip -0.15in
  \caption{DeltaLCA enables designers to rapidly compare the environmental impact (EI) of two PCB designs. In the traditional Life Cycle Assessment (LCA) pipeline (top row), LCA experts manually generate an inventory of parts and match them to a database to calculate the EI. DeltaLCA (bottom row) first automates inventory generation and estimates partial EI for known components, then performs a user-in-the-loop \emph{comparison} using domain-specific heuristics to determine if Design A has a greater EI than Design B.}
  \Description{}
  \label{fig:system_overview}
\end{teaserfigure}


\maketitle

\section{Introduction}
\label{sec:introduction}
For the past 70 years, computing devices and systems have been designed with increasing complexity driven solely by demand for computational power and little foresight to sustainability or disposal. As a result, current estimates of climate warming emissions from the overall information and communication technology (ICT) sector range from 2.1-3.9\%~\cite{freitag_real_2021} of total global emissions and are projected to grow rapidly to 8\% over the next decade if left unchecked~\cite{andrae2015global}. Reducing these emissions requires not only switching to carbon-free energy sources, but also addressing the more specific problem of embodied carbon that comes from device manufacturing. Particularly for ubiquitous consumer devices like smartphones and laptops, manufacturing accounts for over 80\% of life cycle emissions~\cite{apple-device-lca}.

This presents an opportunity to reduce environmental impact (EI) by optimizing future device designs with embodied carbon in mind. Making sustainable design decisions however requires the ability to iteratively analyze a design and determine its carbon footprint which is not possible with today’s tools. The environmental impacts of a device, such as those quoted above, are typically quantified retrospectively through a manual Life Cycle Assessment (LCA) in which a human expert analyzes the impacts of the product’s production, usage, and disposal. This is challenging for computing devices which are complex systems composed of numerous parts including a CPU, memory, power management circuitry, and much more. For example, the printed circuitboard (PCB) of a consumer device may have 500+ parts~\cite{lu_ecoeda_2023}. Moreover, mapping this complex inventory to environmental cost requires data about the semiconductor fabrication processes used to make them which are often proprietary to protect intellectual property. As a result, while there exist tools to support users in computing LCA, performing a rigorous assessment requires significant expertise and could take months for complex designs~\cite{lca-hours1, lca-hours2}, and even experts often lack perfect data. This makes it impractical for designers to compute multiple LCAs for different variations of a design, thereby making it hard to take sustainability into account during the design process itself.


We propose a novel approach to sustainability-focused design for reducing manufacturing carbon foorprint of computing devices at the PCB design stage. Our approach is anchored in two principal ideas: \\
\noindent\textbf{Insight 1:} Our first key insight is that this design tools should be centered around empowering designers with informed decision-making. This approach recognizes that conducting conventional LCAs for each design alternative is not only costly but also impractical as search spaces become larger. We observe however that informed decision-making does not require absolute metrics, but can be done instead with relative comparisons between different designs.  

\noindent\textbf{Insight 2:} Our second key insight is that even if we do not have perfect information, we can leverage domain-specific knowledge of electronics manufacturing to reason about these relative differences between parts or designs.

We combine these insights to develop DeltaLCA, an open source interactive design tool that addresses the dual challenges of automating life cycle inventory generation and data availability by performing comparative analyses for electronics. Using this tool users can input two PCB designs using standard output files from Electronic Design Automation (EDA) tools and the tool will guide the users in determining which one has greater carbon footprint. DeltaLCA leverages electronics-specific LCA datasets and heuristics and tries to automatically rank the two designs, prompting users to provide additional information only when necessary, through a user interface designed to highlight missing data. This approach dramatically simplifies LCA computation by canceling out common components, automatically estimating the carbon footprint of remaining components where possible, and provides a user-friendly interface to input new information or use heuristics to automatically compare the remaining parts.  


Designing a tool that can compare the environmental impact of electronics designs requires addressing two fundamental challenges. The first challenge is that typical LCAs require developing a highly detailed life cycle inventory (LCI) of the raw materials used, their extraction impacts, manufacturing process steps, and the energy they consume. This is particularly challenging in the electronics because every device is a complex combination of integrated circuits (ICs) and other components that are themselves the result of complex semiconductor manufacturing pipelines and global distribution networks. Large components distributors like Digikey have millions of unique components in their catalogs and there are no comprehensive databases of their environmental impact. 


To address this challenge, we develop an automated pipeline to extract relevant LCI data from the outputs of common PCB design tools like KiCAD. We develop a custom Bill-of-Materials (BOM) generator that goes beyond existing tools focused on generating fabrication files to extract detailed information about part classes, footprint sizes, and specifications relevant to environmental analyses using large online parts databases. We then analyze this data to automatically group classes of devices, such as passive components and classes of ICs. We automatically infer essential information for LCA computations such as die size and process technology and combine these with open source measurement data to automate the estimation of carbon footprint for a large class of parts.



While the solutions above provide significant insights into the composition and environmental footprint of a device, there remain significant gaps in the data. This brings us to the second challenge: data availability is a fundamental problem in LCA. To solve this, we look back to our motivation of empowering designers to make sustainable decisions and observe that a common design goal is to reduce environmental impacts compared to a prior version of the same product~\cite{amazon-lca-doc}. In these cases, we care more about the \emph{relative improvement} of the carbon footprint than about the raw impact numbers themselves. As a result, if two designs have a shared set of components with unknown impacts, computing the delta between the designs cancels out these unknowns.

To address the remaining unique parts that are not canceled out, we can take this idea a step further and reduce relative improvement to a binary decision of whether the impact of Design A is greater than Design B. To do this, we leverage both an estimate of the environmental impact (EI) for some parts, whenever these can be computed, and domain-specific heuristics (DSH) which enable us to perform a comparative LCA between parts without having access to full impact data. The final design comparison can then be abstracted into the problem of matching parts of Design A with those from Design B using this partial comparative information. We formulate this matching problem as an integer program, which can be solved efficiently. 
In cases where a complete match is unattainable, we offer a user interface that displays the unmatched parts, allowing users to contribute additional rules for a conclusive comparison.


We summarize our contributions below:
\squishlist
\item DeltaLCA is the first user-in-the-loop design tool that directly empowers electronics designers to make sustainable design decisions by comparing environmental footprints of two different designs.
\item We develop an end-to-end pipeline that directly integrates with common electronics EDA tools to automatically generate a parts inventory and extract the key parameters for LCA computation. 
\item We develop methods to infer proprietary information such as die sizes and device process node and use them to automatically estimate carbon footprint of parts for which information is available. Using these we demonstrate fully automated carbon estimates for two designs with full LCAs and prove our method produces a correct comparison result.
\item We create a set of five domain-specific heuristics for comparing parts for which data is unavailable using proxies such as die size and minimum package area to identify which has a higher environmental impact. In case studies we find that our tool can match an average of 88\% of parts using as few as three heuristics.
\item We develop a matching algorithm that sorts unknown parts into classes and synthesizes the environmental impact estimates and domain-specific heuristics to determine which design is more sustainable. Our algorithm can process inputs with 2500 variables in 0.5~s on a laptop enabling real time comparisons. Our case studies incorporating user input for conclusive comparisons when needed accelerated LCA comparisons to 15~min.
\squishend

\section{Electronics LCA Primer}
\label{sec:lca_primer}

\begin{figure}[t]
  \centering
  \includegraphics[width=\linewidth]{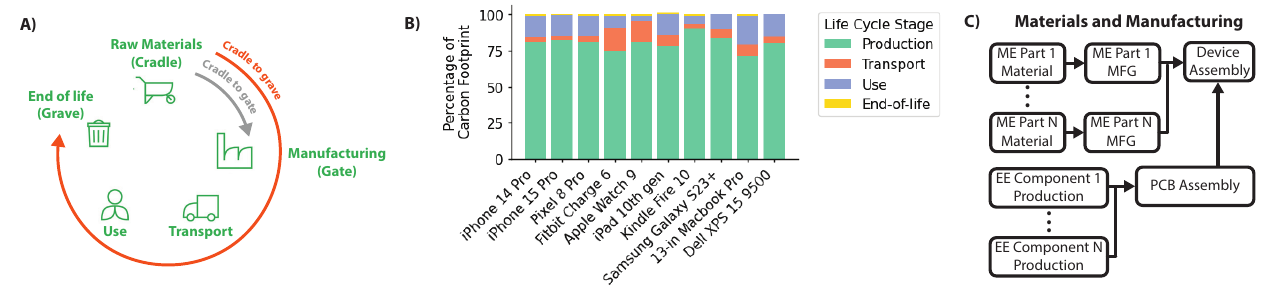}
  \vskip -0.2in
  \caption{A) LCA stages and boundaries. B) Contribution of each stage to total carbon footprint for commercial devices. C) LCA methodology used by Amazon Devices Sustainability~\cite{amazon-lca-doc}.}
  \label{fig:lca-background}
\end{figure}

Performing a Life Cycle Assessment (LCA) involves systematically evaluating the EIs associated with stages of a product's life, from raw material extraction through materials processing, manufacturing, distribution, use, repair and maintenance, and disposal or recycling. This analysis of the total EIs enables manufacturers and consumers to make more informed decisions. The process begins with defining the scope and goal of the assessment, which includes identifying the product to be assessed and the boundaries of the study as shown in \textcolor{blue}{Figure~\ref{fig:lca-background}A}. Common LCA boundaries include ``cradle-to-gate'' assessments which account for raw material extraction through product manufacturing, and ``cradle-to-grave'' which also includes transport to the consumer, usage throughout the product's lifetime, and end of life disposal. The next step is performing an inventory analysis, where data is collected on every input (energy, water, and materials) and output (such as emissions and waste) associated with each stage of the product’s lifecycle. The impact assessment phase interprets this data to understand the EIs, such as global warming potential, ocean acidification, or resource depletion. Finally, the results are interpreted to provide insights and recommendations for reducing the EI.

\textcolor{blue}{Figure~\ref{fig:lca-background}B} compiles information from publicly available product environment reports for a number of recent mobile devices. The results show that for many mobile devices the cradle-to-gate emissions of manufacturing and raw materials comprise 70-80\% of their lifetime carbon footprint and dominate emissions. This is because unlike devices like servers and desktops which are both physically larger and require constant power during their use phase, mobile devices must be portable and designed for energy consumption in mind to enable long battery life. For this class of devices in particular, sustainable design is critical for further emission reduction beyond switching to clean energy sources. As a result, in this work we focus primarily on evaluating carbon footprint in the manufacturing or cradle-to-gate phase.

\noindent\textbf{How is an LCA computed?}
We summarize an example of LCA Methodology published by Amazon Devices Sustainability as a case study of how this process is typically conducted by a domain expert evaluating the cradle to gate segment~\cite{amazon-lca-doc}: A device's bill of materials (“BOM”) is first obtained from an internal Product Lifecycle Management (“PLM”) system. Parts are then assigned to categories as either a mechanical (“ME”) component or process versus an electronic (“EE”) component. Yield losses are accounted for at each level of the BOM from a series of manufacturing and assembly processes based on a multi-level BOM structure. Manufacturing energy is computed by measuring the total energy required for a process and dividing by the number of devices produced to determine unit energy cost. The emissions for each part $E_{part_n}$ in the BOM are summed to determine the total emissions for manufacturing, $E_{MFG} =\sum_{n=1}^{N} E_{part_n}$.

To calculate the emissions, each part must be mapped by an expert to the most appropriate emissions factor available from databases such as ecoinvent, GaBi, industry sources, or academic literature. These emissions factors are generally industry average estimates unless specific supplier data is available. Data availability, however, is a persistent problem in LCA~\cite{bicalho2017lca}. An emissions factor for the IC is chosen based on information available about die size, technology node, and package type. When data is not available, die size is measured for the top five critical ICs in a device and modeled with a conservative emissions factor. Similarly, PCB production is based on total board area, number of layers, and/or mass. Components such as ICs, PCBs, capacitors and resistors are then scaled by mass or area. Scaling by mass is performed by multiplying the component mass $m$ by an emissions factor $ef$ and dividing by the reference mass $m_{ref}$ and loss factor $L$ to account for waste in the production process, $E_{part,EE} = \frac{m \times ef}{m_{ref} L}$.

A similar scaling is applied when using area. The lack of information about IC production presents a key source of uncertainty in this method. The resulting data must then be thoroughly reviewed by experts from the sustainability science team to evaluate data quality issues.

\noindent\textbf{Why is electronics LCA hard?}
While large companies are able to produce these environmental reports for individual high-value products, it is a challenging and time-intensive process for multiple reasons. The first major challenge in electronics LCA is the sheer number of manufactured parts within a device. For example, a device may be composed of over 500 different components across multiple PCBs~\cite{lu_ecoeda_2023}. Within a device like a laptop or smartphone, some parts are themselves sub-assemblies such as an SSD which has multiple ICs on it. These components include an array of different ICs, resistors, capacitors, inductors, and more. Computing a thorough LCA therefore requires evaluating the impacts of each of these individual parts. Second, individual components are themselves manufactured parts that are the result of a highly complex and resource-intensive fabrication pipeline. For example, IC fabrication is a complex multi-step process involving photolithography, etching and deposition processes and more. Third, even within classes of parts such as ICs or capacitors, the manufacturing processes may differ significantly. This makes it difficult to use a single generic process model for each. For example, capacitors use a wide variety of different dielectric materials. Different ICs may use different semiconductor materials optimized for high performance (e.g. GaAs for high frequency RF chips) and certain process technology nodes (often denoted by the minimum feature size, 65 nm, 7 nm, etc.) require fundamentally different technologies with significantly higher environmental costs. 

These fundamental challenges make it very difficult to evaluate LCA during the design phase. This is compounded by the fact that there is a significant disconnect between tools created to perform LCAs and EDA tools for electronics designers. Traditional LCA tools such as GaBi do not have interfaces to interact with PCB design software and operate fundamentally differently. For example, although GaBi has a database of electronic parts, because of the complexity and variety of semiconductor fabrication processes, its models are industry averages not specific to a chip and sorted instead by package type. Additionally, PCB design tools such as KiCAD and EAGLE do not have any way to output the information required to compute sustainablity metrics such as a chip's die size. This combination makes it a highly manual process requiring significant domain expertise to map a component to one's best guess or approximation in a database. In addition to requiring domain expertise, software such as GaBi is proprietary with licensing fees exceeding \$20,000, making it highly inaccessible to electronics designers who seek to optimize for sustainability.

\section{Related Work}


Assessing environmental impacts has been explored in multiple domains. We present a survey of related work on relevant topic areas below.

\subsection{Life Cycle Inventory}
Life cycle assessment (LCA) comprises two phases: Life cycle inventory (LCI) and Environmental impact assessment (EIA). LCI serves as the foundation of LCA \cite{hellweg_emerging_2014}. It involves the compilation and quantification of inputs, outputs, and the potential EIs for a given product throughout its life cycle \cite{suh_methods_2005}. Therefore, the generation of LCI involves comprehensive data collection about all inputs (e.g., raw materials and energy) and outputs (e.g., emissions and waste) associated with a product's life cycle, which is often the most costly and time-consuming phase~\cite{wernet_ecoinvent_2016}. Different approaches for generating LCI have been explored for decades with two main focuses: accuracy and boundary completeness \cite{islam_review_2016}. Process-based LCI examines every process involved in the life cycle, often used in the studies of products, such as energy products \cite{guezuraga_life_2012} and electronic products \cite{andrae_life-cycle_2016, smith_life_2018}; input-output-based LCI connects the economic IO model to LCA, offering a broader perspective \cite{leontief_environmental_1970}; hybrid approach combines elements of these two, offers greater accuracy and boundary completeness at the cost of greater time and complexity. Despite decades of research, however, the generation of LCI remains complex and time-consuming and requires a significant amount of manual work.

\subsection{Environmental Impact Assessment for ICT}
A number of works have attempted to estimate the overall impact of the ICT industry as a whole~\cite{freitag_real_2021,andrae2015global}. Additionally others have developed tools to model embodied and operational emissions, corresponding to the production and use stages of computing devices. The dominating source of emissions, as discussed by Gupta et al. \cite{gupta_act_2022, gupta_chasing_2020}, is shifting from operational activities toward hardware production for many devices. 

Computing devices typically consist of numerous and varied components, each class of which includes a diverse range of materials and processes involved in the production which complicates LCA computation \cite{djekic_scientific_2019}. In response to these challenges, there has been a growing body of work focused on developing modeling tools. Much of this work, however, has been concentrated on larger-scale applications, such as data centers \cite{elgamal_carbon-efficient_2023, wang_peeling_2023, acun_carbon_2023}, or IC level designs~\cite{gupta_act_2022}. In this work we seek to complement these works by developing a tool for circuitboard level designs to empower designers in the ubiquitous and mobile computing communities to consider sustainability metrics.

\subsection{Comparative Life Cycle Assessment}\label{sec:clca}
Prior work on sustainable development has explored the idea of performing comparative LCA  as a decision-making tool, enabling stakeholders to evaluate and compare the EIs of various alternatives and successors \cite{zhang_comparative_2022, liu_future_2014}. Comparative LCA is also crucial for making public comparative claims. Despite its importance, these works have only done this by evaluating two complete LCAs and comparing the resulting outputs. This approach to comparative LCAs presents several challenges. First, the cost and time associated with full LCA studies are often prohibitive, particularly in industries like electronics characterized by rapid product development cycles. Second, the complexity of modern supply chains, especially in sectors like electronics, makes the quantification of all processes a time-consuming and resource-intensive task \cite{lca-hours1,lca-hours2,amazon-lca-doc}. Third, the requirement for extensive and often confidential data makes full LCAs daunting. 

In response to these challenges, recent work developed various methods to streamline the LCA process, i.e., simplified or screening LCA. For instance, certain analysis, known as cradle-to-gate, gate-to-gate, etc, ignores certain upstream or downstream processes \cite{gradin_common_2021}. Others focus on a narrower range of evaluated environmental metrics \cite{bala_simplified_2010}, which streamlines the inventory phase by reducing the inventory parameter considered in the selected categories. Other approaches simplify the analysis by using mass as a coarse-grained indicator of raw material used \cite{nissen_comparison_1997, kaebemick_simplified_2003} to perform quicker assessments. 

Furthermore, traditional LCA is susceptible to variability in LCA methodologies and software tools between studies. Inconsistencies in results can be attributed to differences in:
\begin{itemize}
    \item \textbf{System boundaries.} System boundaries determine which processes and life cycle stages are included in the assessment. Even one stage difference can lead to significant variations in the results.
    \item \textbf{Database differences.} If two studies use different data sources, for example, choosing thermal power and nuclear power electricity within the same region could cause 164x of difference \cite{ding_comparative_2017}. Variations across states and even countries can also cause large differences in the calculated LCA results.
    \item \textbf{Methodological differences.} LCA methods can vary in terms of their assumptions. Differences in factors like allocation methods, emissions factors, and modeling assumptions can lead to variations in results.
\end{itemize}

This further complicates the LCA comparison between different designs. We propose a different approach that enables decision-making through relative comparisons between different stages of designs. Canceling out common components between design iterations dramatically simplifies the number of parts to evaluate, and we can then develop ways to reason about which components have more environmental impact.


\subsection{Sustainability in HCI and Ubicomp}
The field of HCI and UbiComp has recently seen an increased interest in sustainable computing. This trend aligns with the broader societal shift towards environmental consciousness. Previous research has focused on unmaking \cite{song_unmaking_2021, cheng_functional_2023}, the development of biodegradable \cite{arroyos_tale_2022, cheng_swellsense_2023, vasquez_myco-accessories_2019} and recyclable materials \cite{zhang_recyclable_2023} for substrates, and innovative design tools that facilitate the reuse of electronic components \cite{lu_ecoeda_2023}. It is worth noting that \textit{EcoEDA}~\cite{lu_ecoeda_2023} aims to extend the lifespan of electronic parts, thus mitigating the EI of electronic waste. 

Most existing research in sustainable computing has been directed towards the end-of-life stage of electronics and primarily focused on new materials. While novel materials are an important part of a holistic sustainability solutions, the EI of a PCB is also significantly influenced by decisions made during its design stage, including the choice of ICs, size, and the integration level of components. Kaebemick et al.~\cite{kaebemick_simplified_2003} emphasize the importance of incorporating environmental considerations into the early design stages. Our work aims to fill the gap in tools to evaluate EI and empower the computing research community to integrate sustainability metrics into the PCB design stage.

\section{System Overview}

\label{sec:system_overview}
This section presents an overview of DeltaLCA, a system to assess the relative EI between two input designs $\mathcal{A}$ and $\mathcal{B}$.
The differences between the traditional way of comparing two designs and DeltaLCA are illustrated in \textcolor{blue}{Figure~\ref{fig:system_overview}}.

\paragraph{Traditional LCA workflow}
As described in Section~\ref{sec:lca_primer}, a traditional LCA is performed in two phases, the inventory phase and the assessment phase (green and orange boxes), respectively, in \textcolor{blue}{Figure~\ref{fig:system_overview}}.
Both phases are entirely performed by LCA experts. 
In the inventory phase, the expert lists all relevant parts of the design file to establish a Complete Parts List.
Then, for each part in that list, the expert needs to fill in EI data from an LCA database, such as ecoinvent \cite{wernet_ecoinvent_2016}. 
Parts, especially recent electronic parts, are often not readily available in these databases, and the expert often needs to search for the best approximation using their domain knowledge, making it impossible to automate this process. 
In the assessment phase, the expert imports additional processes and establishes a flow diagram in an LCA software, such as GaBi.
Creating a flow diagram requires an understanding of both the LCA-specific software and knowledge about the secondary processes and resources needed to manufacture parts.
This entire process is done for both input designs, resulting in two EI results.
Finally, once the expert has these numbers, they can compare the two designs.

\begin{figure}[t]
  \centering
  \includegraphics[width=0.6\linewidth]{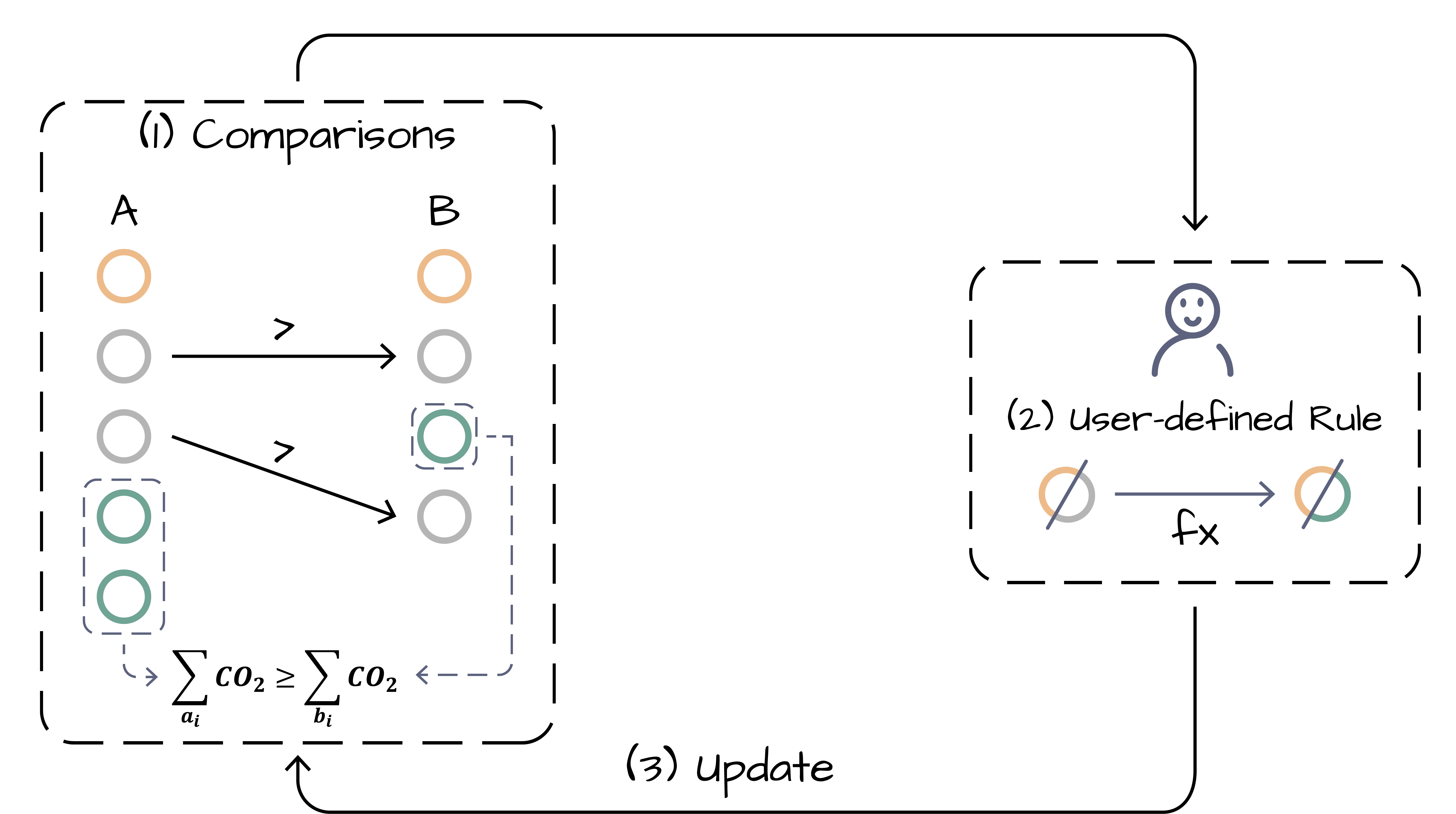}
  \caption{Our comparison algorithm uses parts either for pairwise comparisons (grey nodes) via heuristics (edges) or for carbon footprint comparison (green nodes). (1) The initial comparison results are first delivered to the user. (2) Drawing from the user's domain knowledge, the user adds comparison rules using any parts available to solve the unmatched parts (orange nodes). (3) The results can be updated, taking into account the user-defined rules.}
  \label{fig:user_loop}
\end{figure}

\paragraph{DeltaLCA workflow}
In DeltaLCA, we completely automate the inventory phase. We then replace the two assessment phases with a single \emph{comparative LCA phase}, which requires minimal user feedback. 

For the inventory phase, we make use of both insights from Section~\ref{sec:introduction} to automatically establish a Complete Parts List and a Partial LCI.
Focusing on a domain-specific tool, in our case a tool for electronic designs, we can leverage our domain knowledge to build an LCA-targeted parser of design files (Insight 2).
For each part, we automatically parse all the information available in public datasets and auxiliary sources.
The result of this step is a Partial LCI (see Section~\ref{sec:lci}).

A Partial LCI contains three kinds of parts:
\begin{enumerate}
    \item Parts which are identical and which are present in both $\mathcal{A}$ and $\mathcal{B}$. 
    We can cancel these parts out before the comparison algorithm.
    \item Unique parts for which we have sufficient LCA information to compute their EIs.
    \item Unique parts for which we have only partial information.
\end{enumerate}

Parts from type $(1)$ are removed during pre-processing, whereas parts from type $(2)$ are usually encountered in a traditional LCA.
To compare parts with partial information $(3)$, we use expert-designed \emph{heuristics}, which allow us to determine that part $a_i$ from $\mathcal{A}$ will have greater EI than a part $b_j$ from $\mathcal{B}$ without ever computing the actual EI of either part. 
Intuitively, these heuristics can be thought of as conservative rules, e.g. a heuristic may state that if two ICs are manufactured with the same processes but one is significantly larger than the other, the larger one will have greater EI. 
This means that the key algorithmic challenge for reasoning about parts with partial information is to perform a \emph{pairwise matching} between two sets of parts using these heuristics. 
Importantly, this matching algorithm must also optimally leverage the information that enables computing full EIs for some of the parts. 
We tackle this challenge with a comparison algorithm which is explained in Section~\ref{sec:comparison}.

\paragraph{User-in-the-loop}
The output of the comparison algorithm is two design subsets $\mathcal{A_\delta}$ and $\mathcal{B_\delta}$ (nodes except for the orange ones in \textcolor{blue}{Figure~\ref{fig:user_loop}}) for which we can show that the EI of the former is bigger than the EI of latter.
For the remaining parts (orange nodes), i.e. $\mathcal{A}\setminus \mathcal{A_\delta}$ and $\mathcal{B}\setminus \mathcal{B_\delta}$, our algorithm either has insufficient information about the parts, or design $\mathcal{A}$ is simply not more environmentally impactful than design $\mathcal{B}$.
In both cases, the user is presented with the comparison result and the remaining parts.
To refine the assessment, they can then provide more information to the system by completing part information that the inventory phase has been unable to retrieve or by providing additional, user-defined rules about the relative impact between parts.
Once more information has been provided, either about the parts themselves or about the relationship between parts, the comparison result can be updated.
Our interactive feedback loop is illustrated in \textcolor{blue}{Figure \ref{fig:user_loop}}: (1) the two designs are automatically compared, and the results are delivered to the user; (2) the user can provide additional information to refine the comparison; (3) the comparison result gets updated, taking into account the new information.

In summary, DeltaLCA automates the inventory phase and the assessment phase as much as possible by leveraging domain-specific knowledge and by canceling out uncertainties through part comparisons.
As opposed to the traditional LCA comparison, with DeltaLCA the user is only needed at the end for the actual comparison and not throughout the entire process.

\section{Automated Life Cycle Inventory}\label{sec:lci}

The first step in our DeltaLCA pipeline is generating an LCI from PCB design files. Traditional methods for LCI require time and expertise for two key reasons. First, the component models available in LCA software databases do not have a one-to-one mapping to the parts on a PCB which prevents automation. Second, data for many parts is unavailable. 


To achieve the desired level of accuracy and automation, we implemented a four-phase pipeline: 1) parsing the raw design files locally. The goal of this step is to extract basic information about the components and board configuration within the PCB design relevant to LCA. 2) interfacing with external datasources to enrich our resulting part lists with up-to-date attributes of each component. 3) inferring core specifications, typically kept confidential by manufacturers, with all available information. These specifications, such as the die area of an IC, are crucial for assessing the EIs of ICs. 4) estimating the carbon footprints of standard components with complete information. We describe each phase in detail below.

\begin{figure}[b]
  \centering
  \includegraphics[width=\linewidth]{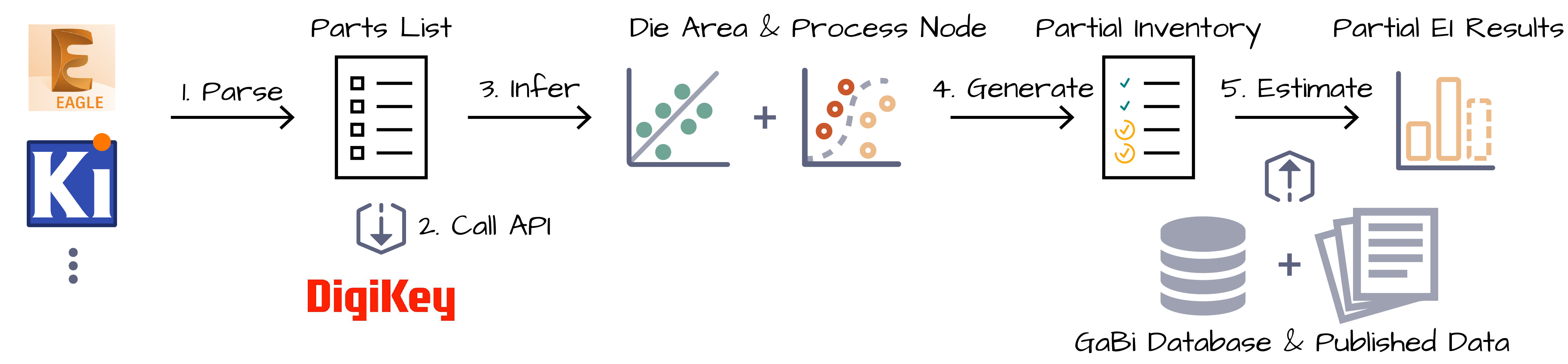}
  \caption{Our automatic LCI pipeline. We start by parsing the PCB design files from common design software into a parts list, then infer core specifications like die sizes leveraging online resources, and finally generate the partial inventory with partial EIs based on publicly available data.
  }
  \label{fig:lci_pipeline}
\end{figure}

\subsection{Create Parts List}
Generating parts list from raw PCB design files and accurately categorizing components, while simultaneously extracting detailed information such as size and series, is challenging. We note that computing LCA requires information beyond a typical BOM which is designed for purchasing parts and fabrication and does not contain information about device size or specifications. Generating an accurate parts list is inherently complex due to several key factors: 1) even when PCB designs are created using the same design software, such as Eagle, the components library can significantly differ between projects, resulting in diverse and often customized component symbols and footprints. This inherent variability makes it challenging to establish a uniform and robust parsing algorithm. 2) The physical packaging of electronic components can differ significantly, making it challenging to employ a one-size-fits-all algorithm for component categorization. Moreover, classifying between similar components, such as different passive components (e.g., resistors, capacitors, inductors) versus IC packages, necessitates a high degree of domain-specific knowledge. 3) Component information is scattered throughout the design files, requiring sophisticated data extraction pipelines to gather the relevant data. 4) Detailed electrical specifications of components, such as power consumption, are often not included in the PCB design files. This lack of comprehensive data necessitates the integration of external databases. We focus on three primary EI sources: ICs, PCB substrates, and other electronic components in our implementation.

\subsubsection{Parse Raw PCB Design File}
Our automation pipeline is implemented in Python, starting with the extraction of pertinent information from the raw PCB layout files. The parsing method involves several critical steps. First, we calculate the board's dimensions by identifying and processing the coordinates of the outermost wires in the PCB plane. Then, we assess the number of layers by enumerating the activated route layers between the \textit{Top} (which contains the copper on the top of the board) and \textit{Bottom} layer markers, indicative of a multilayer PCB structure. The core functionality of parsing begins with systematically iterating through each element in the PCB design, and the goal is to sort all electronic components into four main categories: passive components (e.g., resistor, capacitor, inductor), active components (e.g., diode, transistor), ICs (e.g., microcontroller), and misc (e.g. connectors).

Drawing on research in electronic components guidelines and standard topologies, we propose the following heuristics for the initial categorization:

\begin{itemize}

\item \textbf{Element name prefix.} We can effectively categorize SMD resistors, capacitors, and inductors into their respective classes by recognizing the prefixes `R', `C', and `L', respectively. These are followed by numeric size identifiers such as `0402' or `0603', which signify the length and width of the package. We employ a set of regular expressions that scan for characteristic prefixes in the designators of components in the raw PCB layout.

\item \textbf{Footprint topology.} The physical layout and pin configuration of components also provide significant clues. Specific topology and pad positions are often unique to certain component types. For instance, discrete surface mount transistors are commonly associated with three pins where the three pads in the PCB layout may form a triangular shape; operational amplifiers (op-amps) are often available in 8-pin DIP or SOP packages and have symmetrical pin layouts; microcontrollers are characterized by a higher pin count, typically having 16 pins or more.

\item \textbf{Component correlation.} We analyze the network of connections to further aid in their identification. Microcontrollers, which usually serve as the main processing unit on a PCB, can be differentiated by examining the highest density of connections to their footprints. This allows us to distinguish between various ICs and other components that cannot be classified solely based on their element names.

\end{itemize}

We also determine the core package size of each IC by analyzing the \textit{tDocu} or \textit{tPlace} layer, i.e., the silkscreen. We calculate the extremities of the wires that define the positioning of components in the PCB layout. We exclude components such as fiducial holes and screw terminals from the parsing process to maintain a focus on electronic components. All parsed information is stored with a set of attributes, includingcomponent name, package type, package area, and a quantity which is incremented for each duplicate component.

\subsubsection{Call External API}
For each IC, we then initiate a keyword search via Digikey's Product Information API, with the part name serving as the search query. Integrating real-time API calls to comprehensive online databases of electronic component distributors like Digikey enriches the parts list with additional up-to-date technical attributes, such as operational frequency and number of GPIO, that are not typically included within the layout design files themselves. This also enables cross-validation for the information extracted from the user files to resolve custom component libraries.

In the case of empty returns, we attempt modified fuzzy searches by stripping off the suffix to accommodate different naming conventions and improve the search completion rate. Upon successfully receiving a response, DeltaLCA parses the detailed attributes of the top-matched search result returned. We extract the key attributes of ICs such as supplier device package, memory size, and number of pins from the product's attributes. These attributes will be used to infer the specifications of the die, which will be discussed in the section below. Additionally, a fuzzy string matching algorithm is applied to accurately map and retrieve product attributes that may be represented by various terminologies across different data entries.

In \textcolor{blue}{Table~\ref{tab:BOM_comparison}}, we compare our method against a range of existing BOM generation tools. The baselines include the built-in BOM generators within commonly used PCB design software such as EAGLE and KiCAD, as well as popular open-source tools like KiCost \cite{kicost} which can extract component data from several distributors' web servers and InteractiveHtmlBom \cite{InteractiveHtmlBom} which has robust local file parsing capabilities. Our parser is the only one that extracts all of the available part information needed to compute LCA.

\begin{table}[h]
  \centering
  \includegraphics[width=0.9\linewidth]{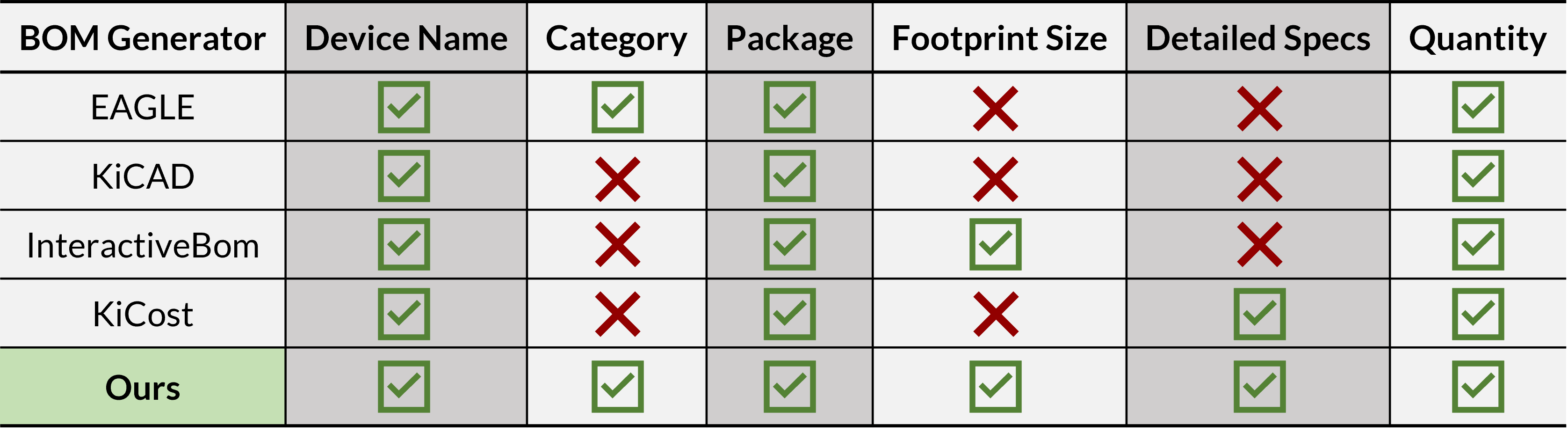}
  \caption{The comprehensiveness of our method compared to existing BOM generation tools.}
  \label{tab:BOM_comparison}
\end{table}


\subsection{Generate Partial Inventory}
Having established a robust pipeline for local design file extraction with external API integration, we next focus on inferring die specifications in ICs and then generating a partial inventory. 

Integrated Circuits, such as microcontrollers, represent the most substantial portion of the environmental footprint in the device manufacturing stage and are more significant than passive components like resistors and capacitors or the PCB substrate itself~\cite{zhang_recyclable_2023, arroyos_tale_2022}. This predominance stems from the high energy cost for semiconductor fabrication plants and raw material cost during the complicated fabrication processes~\cite{gupta_act_2022}. 

The die size and process technology node of IC are two main indicators of the environmental footprint from the manufacturing stage because 1) die size determines the amount of raw material used; 2) process technology node which determines the complexity of fabrication: the lithography equipment, patterning process, and associated energy consumed during fabrication. More advanced (smaller) technology nodes and larger die sizes also typically lead to lower yields, which further increase the fraction of the fabrication energy and resources used per chip. Consequently, inferring the die size and process technology node of ICs within the desired error range is crucial for estimating the environmental footprint of a PCB design.

\begin{figure}[b]
  \centering
  \includegraphics[width=\linewidth]{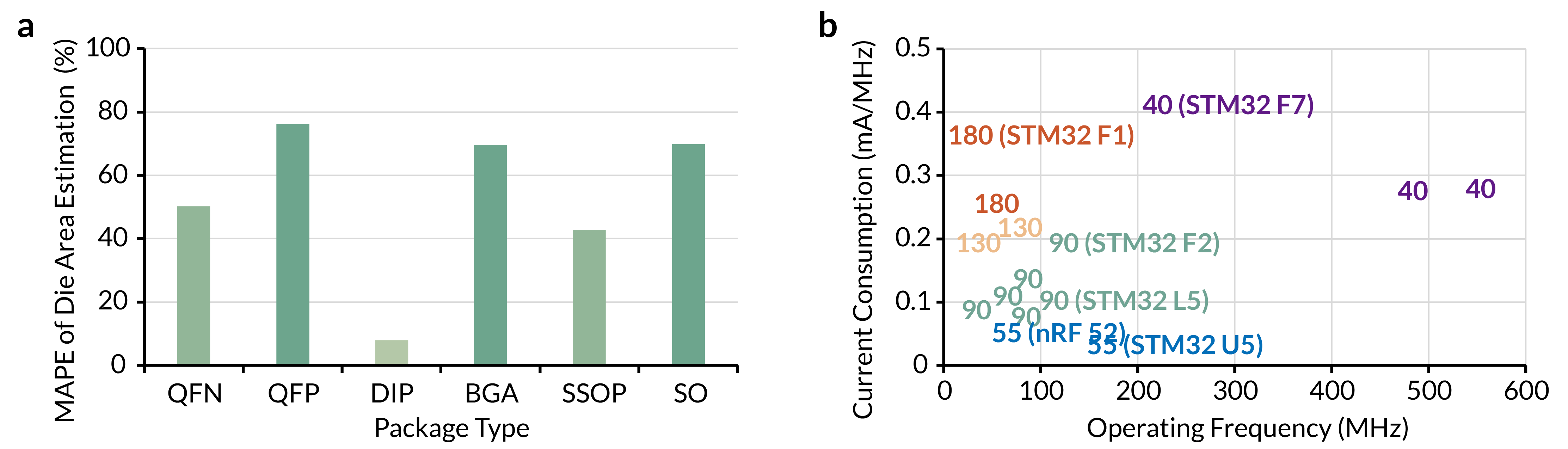}
  \caption{(a) Performance by package type of our die size estimation in Mean Absolute Percentage Error (MAPE) (N >= 3). QFN: Quad Flat No-lead; QFP: Quad Flat Package; DIP: Dual In-line Package; BGA: Ball Grid Array; SSOP: Shrink Small Outline Package; SO: Small Outline. (b) Trends of process technology nodes across 15 popular MCU series. Each data point represents an MCU series, labeled with its process node in nm, plotted against its maximum operational frequency (MHz) and active current consumption normalized by frequency (mA/MHz).
  }
  \Description{}
  \label{fig:die_specs}
\end{figure}

\subsubsection{Infer Die Area}
In practice, calculating the precise die size of an IC often requires access to the original IC's design files from semiconductor packaging suppliers. However, these types of proprietary information are typically not readily disclosed by design houses, posing a significant barrier. 


We aim to develop a generalized alternative method for approximating die size with readily available information: the dimension of the IC package and package type. The dimensions of the package and its structural characteristics deterimine the maximum silicon area it can house. For example, QFP packages usually have a traditional lead frame with leads extending from all four sides of the package, requiring space for the metal contacts; QFNs do not have leads protruding from the package, and instead, they have an exposed metal pad on the bottom surface; WLCSPs integrate leads within the chip fabrication process, making these a close approximation of the bare silicon die size. 

Using the GaBi Electronics Extension dataset, we compile a table of die area coefficients, which represent the percentage of  each package type occupied by actual silicon. This enables us to rapidly estimate the die area directly from the package size and type with a sufficient level of accuracy. 

Using a dictionary of specific package types and their corresponding die size coefficients, we scale each package size by its coefficent to obtain estimated die area. For packages with only one dimension available, the package is assumed to be square, and the area is calculated accordingly. A fuzzy string matching algorithm is applied to determine the best match for the package type within our coefficient dictionary, accounting for any variations in naming conventions and missing package info in the dictionary.

To evaluate our method, we used the Mean Absolute Percentage Error (MAPE), a normalized error metric, as our primary measure. This choice is relevant because the die sizes differ across different MCU series, resulting in varying scales of ground truth values. As shown in \textcolor{blue}{Figure~\ref{fig:die_specs}a}, our method demonstrates MAPE of 50.20\%, 76.22\%, 7.93\%, 69.65\%, 42.81\%, 69.83\% for QFN, QFP, DIP, BGA, SSOP, and SO packages, respectively. These values indicate the relative deviation of our estimated die areas from the actual values. Notably, more complex packages like QFP and BGA exhibit higher indeterminacy due to their complex internal structures. 


\subsubsection{Infer Process Technology Node}
To infer the process technology node used in an MCU, we establish a correlation between the maximum operational frequency and the active current consumption. As the process node size decreases, which implies smaller transistor dimensions, the gate capacitance is reduced. This reduction is proportional to the transistor's switching time due to the diminishing resistive-capacitive (RC) delay, and lower power consumption by minimizing the power dissipation through the channel \cite{shahidi_chip_2019}. Consequently, smaller process nodes enable higher operational frequencies while concurrently decreasing active current.

Such dual observations act as a heuristic to infer the process technology node. For instance, an MCU operating at a higher frequency with a relatively lower active current than another MCU with similar functionality indicates the use of a more advanced process node. \textcolor{blue}{Figure~\ref{fig:die_specs}b} shows the process technology node for 15 commonly used MCU series with publicly available data in relation to their operational frequency and corresponding active current consumption normalized by frequency. We use this data to construct a lightweight classification model to estimate the process node by fitting the known specifications of an MCU with established trends in semiconductors. This method leverages empirical data from semiconductor technology advancements to provide a non-invasive estimation technique that does not rely on direct semiconductor foundry data.

\subsection{Estimate Partial Environmental Impact}
\label{sec:precompute_fixed_components}
To facilitate a standardized comparison of the environmental footprint in PCB design, we first estimate the cradle-to-gate carbon footprints of standard fixed components such as resistors, capacitors, inductors, and substrates, that are commonly present in every PCB, in CO2 equivalent. We aim to streamline the process, particularly for users who are new to LCA.

Despite the variability in the materials used for common passive components (capacitors can be made from multilayer ceramic, tantalum, or electrolytic dielectric layers) prior work has shown their EIs are relatively similar \cite{smith_life_2018}. Furthermore, compared to major footprint sources like PCB substrates and ICs, the environmental footprint of these passive components is small. Consequently, assigning a single LCA value per component does not result in significant errors. 

In our assessment, we standardize the LCA values for these components based on their package size and the results are shown in \textcolor{blue}{Table~\ref{tab:LCA_values_rcl}}. The baseline values are derived from previous research \cite{smith_life_2018} with a functional unit of 1 kg for each component type (e.g., 1 kg MLCC capacitors equates to 97 kg of CO2-eq). For resistors, the environmental footprint ranges from 0.01 grams of CO2 equivalent per piece for 0201 package to 0.6 grams for 1206. Capacitors follow a similar trend, where the EI increases with package size, starting at 0.036 grams of CO2 equivalent for the 0201 size, and rising to 1.067 grams for 0805. Inductors range from 0.022 grams of CO2 equivalent for 0201 to 1.358 grams for 0805. For the PCB substrate material, we selected FR-4, the most commonly used PCB substrate material, which has a standardized environmental footprint of 0.006125 grams of CO2 equivalent per square millimeter per 1mm-thick layer \cite{liu_future_2014, ozkan_life_2018, zhang_recyclable_2023}. For ICs with complete information including die area and process node, we also estimate their carbon footprints using the aggregate carbon footprints per area from Gupta et al.'s work \cite{gupta_act_2022}.


\begin{table}[h]
\centering
\begin{tabular}{@{}llllll@{}}
\toprule
Package (Imperial)  & Resistor  & Capacitor & Inductor \\ \midrule
0201                & 0.010     & 0.036     & 0.022 \\
0402                & 0.040     & 0.146     & 0.078 \\
0603                & 0.120     & 0.611     & 0.330 \\ 
0805                & 0.600     & 1.067     & 1.358 \\\bottomrule
\end{tabular}
\vspace{.15cm}
\caption{Standardized environmental impacts in gram CO2-eq for common passive electronic components (resistors, capacitors, and inductors).}
\label{tab:LCA_values_rcl}
\end{table}
\section{Comparative Impact Assessment}\label{sec:comparison}
\subsection{Overview}


The comparison algorithm takes the partial LCIs from the inventory phase for two designs $\mathcal{A}$ and $\mathcal{B}$.
Without loss of generality, we assume that we want to show that the environmental impact, e.g. carbon footprint, of $\mathcal{A}$ is \emph{bigger} than $\mathcal{B}$.
In general, our algorithm will be able to show this for two subsets $\mathcal{A}_\delta$ and $\mathcal{B}_\delta$. 
Our primary goal is to design an algorithm where $\mathcal{B}_\delta$ is maximized, thereby enhancing the likelihood of demonstrating that $\mathcal{B}$ has the lowest environmental impact with minimal or no additional user input (i.e., when $\mathcal{B}_\delta \equiv \mathcal{B}$).

As mentioned in Section~\ref{sec:system_overview}, at this stage, partial LCIs contain parts with complete LCA information for which we can perform direct carbon footprint comparison and parts with only partial information which will be compared using pairwise heuristics.
But how do we combine heuristic-based comparisons with carbon footprint based comparisons? One straightforward approach would be to first perform an LCA comparison for parts with complete information, so parts that have carbon footprint information, and then, in a second stage to perform a pairwise comparison based on heuristics.
However, this could easily lead to selection problems. Suppose $\mathcal{A}$ has only parts with complete information and $\mathcal{B}$ has parts with partial information. The first step would consume all parts of $\mathcal{A}$, even if not all of the per-part carbon footprints are necessary to weigh up against the carbon footprint of $\mathcal{B}$. Then, $\mathcal{A}$ would not have any parts left for the pairwise comparison.
From this example, we can see that the selection of parts used for the carbon footprint comparison and the pairwise, heuristic-based comparison are interdependent. 
We formulate this interdependent selection problem as an integer program, as described in Section~\ref{sec:integer_program}.
Finally, we discuss our implementation and evaluate the performance in Section~\ref{sec:integer_program_impl}.


\subsection{Heuristics}
\label{sec:heuristics}

We define a \emph{heuristic} $\mathbf{H}$ as a map from a subset of parts of design $\mathcal{A}$ and a subset of parts of design $\mathcal{B}$ to a decision about which subset has a higher carbon footprint, or if neither of them subsumes the other.
The heuristic decision is based on certain part attributes and we only assign heuristics between parts that have the necessary attributes available.
Heuristics are based on domain knowledge of the application domain and example heuristics to compare PCB components are listed below. We note that these can be extended and modified by the user based on available information:

\begin{enumerate}
\item \textbf{Package size for the same type of passive components.} For passive components, such as capacitors and resistors, those of identical dimensions consume comparable amounts of raw materials, making their environmental impact comparable, irrespective of variations in material composition.

\item \textbf{Chips with the same core and architecture.} Chips that share a common standard core and architecture (e.g., ARM Cortex-M), are considered equivalent in our analysis if they differ only in packaging type. This is underpinned by two factors: the environmental impact of the package is relatively overshadowed by the silicon part; a shared core impies identical manufacturing processes and material usage, thereby equating their environmental footprints.

\item \textbf{Equivalent process.} In cases where components are analogous in dimensions and weight, we assume that their transportation-related environmental costs are comparable. Furthermore, for components of the same category, the electricity cost in the assembly process, such as soldering, is consistent. 

\item \textbf{Chip die size.} The impact of chip die sizes is twofold. Larger dies not only suggest increased raw material consumption but also tend to result in lower production yields, both contributing to a larger environmental impact.

\item \textbf{Process technology node.} More advanced (smaller) process technology node requires more steps and complexity in fabrication. This includes the need for advanced lithography equipment and multiple patterning processes, culminating in elevated energy consumption.

\item \textbf{Diode size.} Across various diode types, such as photodiodes, LEDs, and Schottky diodes, size is an effective indicator of raw material consumption. Given their similar semiconductor structures, the size of the diode correlates directly with the environmental costs associated with raw materials.

\end{enumerate}

We use the first three heuristics to identify parts in $\mathcal{A}$ and $\mathcal{B}$ which are "identical", i.e. which parts we assume to amount to the same carbon footprint, given the partial information available to use.
These parts are being removed before the comparison algorithm.



Note that we have no means of resolving conflicting heuristics.
For example if between a chip $a_i$ and a chip $b_j$, $a_i$ has a bigger die size (heuristic 4), but $b_j$ has a smaller process node (heuristic 5), then we have two conflicting pieces of information.
Without further information about how to compare these two chips, we have no means of reasoning about them.
We therefore remove all heuristics between parts which have conflicting heuristics before our comparison algorithm.


\normalsize
\subsection{Comparison Algorithm}
\label{sec:integer_program}
\begin{figure}[t]
  \centering
  \includegraphics[width=0.5\linewidth]{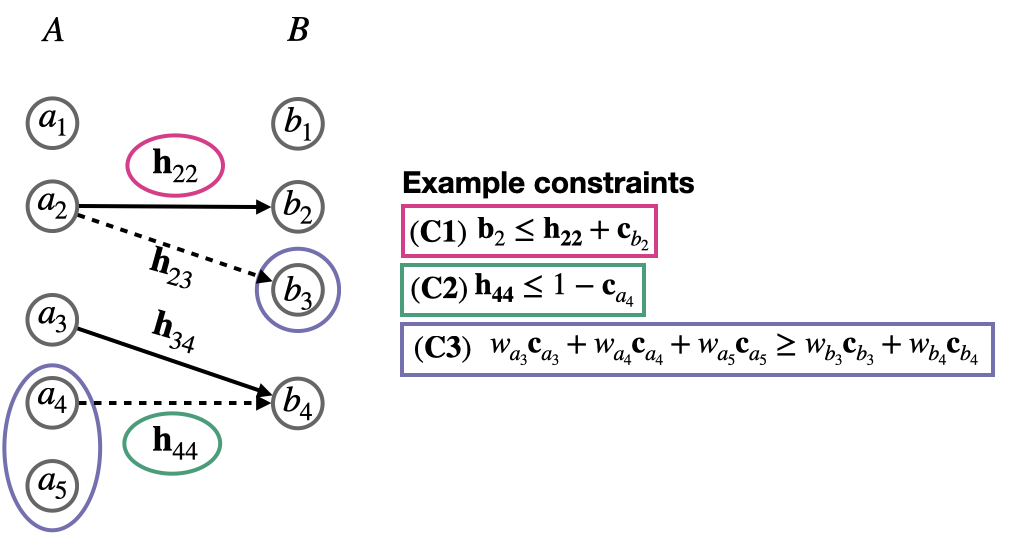}
  \caption{Input to this example problem is the bipartite graph, including the heuristic edges and the carbon footprint weights, see \textcolor{blue}{Table~\ref{tab:example_bip}}. 
  The three example constraints illustrate the three constraints from our integer program formulation.}
  \Description{}
  \label{fig:example_bip}
\end{figure}

In the following, we explain the general integer program formulation for the comparison of two designs.
Please refer to \textcolor{blue}{Figure~\ref{fig:example_bip}} for a toy example.

\begin{table}[t]
\centering
\begin{tabular}{|p{15mm}|p{15mm}|}
\toprule
Carbon footprints $\mathcal{A}$  & Carbon footprint \newline variables $\mathcal{A}$\\ \midrule
$w_{a_1}=N/A$                & $\mathbf{c}_{a_1}=0$ \\
$w_{a_2}=N/A$                & $\mathbf{c}_{a_2}=0$ \\
$w_{a_3}=10$                & $\mathbf{c}_{a_3}=0$ \\ 
$w_{a_4}=10$                & $\mathbf{c}_{a_4}=1$ \\ 
$w_{a_5}=10$                & $\mathbf{c}_{a_5}=1$ \\\bottomrule
\end{tabular}
\begin{tabular}{|p{15mm}|p{15mm}|p{15mm}|p{15mm}|}
\toprule
Carbon footprints $\mathcal{B}$  & Carbon footprint \newline variables $\mathcal{B}$ & Part \newline variables $\mathcal{B}$ & Heuristic variables\\ \midrule
$w_{b_1}=N/A$                & $\mathbf{c}_{b_1}=0$& $\mathbf{b_1}=0$ & $\mathbf{h_{22}}=1$\\
$w_{b_2}=N/A$                & $\mathbf{c}_{b_2}=0$& $\mathbf{b_2}=1$ & $\mathbf{h_{23}}=0$\\
$w_{b_3}=10$                & $\mathbf{c}_{b_3}=1$& $\mathbf{b_3}=1$ & $\mathbf{h_{34}}=1$\\ 
$w_{b_4}=10$                & $\mathbf{c}_{b_4}=0$& $\mathbf{b_4}=1$ & $\mathbf{h_{44}}=0$\\\bottomrule
\end{tabular}
\caption{Carbon footprint values and variable assignments for the example problem from \textcolor{blue}{Figure~\ref{fig:example_bip}}. Parts for which we do not have complete LCA information have no available carbon footprint. The binary variables are assigned such that the objective function is maximized.}
\label{tab:example_bip}
\end{table}

\paragraph{Objective function} The goal of the selection problem is to select parts of $\mathcal{A}$ and parts of $\mathcal{B}$ such that each part of $\mathcal{B}$ is shown to be of lesser carbon footprint than some part of $\mathcal{A}$, either by means of direct carbon footprint comparison or by selecting pairwise heuristics.
More specifically, we formulate this selection problem as a binary integer program where we associate to each part $\mathbf{B}_i$ a binary variable $\mathbf{b}_i \in \{0, 1\}$. 
$\mathbf{b}_i$ is equal to $1$ if it the part $\mathbf{B}_i$ has been included in the LCA comparison.
The objective function is shown in Eq.\ref{eq:obj_function}. We are maximizing the number of compared parts of $\mathcal{B}$ since we want to do a comparison for as many parts as possible of that design.
\begin{equation}
    \max \sum_i \mathbf{b}_i
\label{eq:obj_function}
\end{equation}

\paragraph{Constraints} $\mathbf{b}_i$ should only be put to $1$ if it has been taken into account by a pairwise heuristic comparison or by means of its carbon footprint, which we translate via the constraint in Eq.\ref{eq:constraint_part_b}. 

\begin{equation}
(\mathbf{C1}):    \mathbf{b}_i \le \left(\sum_{\mathbf{h} \in \mathcal{H}_{\mathbf{b}_i}} \mathbf{h}\right) + \mathbf{c}_{\mathbf{b}_i}
\label{eq:constraint_part_b}
\end{equation}
where $\mathcal{H}_{\mathbf{b}_i}$ is the set of all binary heuristic variables for heuristics which show that the part $\mathcal{B}_i$ has been subsumed by some part of $\mathcal{A}$ and $\mathbf{c}_{\mathbf{b}_i}$ is a binary variable which is equal to $1$ if $\mathcal{B}_i$ has been used in the carbon footprint computation.
In the toy example in \textcolor{blue}{Figure~\ref{fig:example_bip}}, $\mathbf{b}_2$ has been successfully put to $1$ because the heuristic $h_{22}$ has been selected and $b_2$'s carbon footprint has not been used.

As mentioned earlier, each part $\mathcal{A}_i$ of $\mathcal{A}$ can only be used in one heuristic. But if an $\mathcal{A}_i$ has already been used in the carbon footprint computation, it cannot be used for pairwise heuristic comparison. This is enforced by the constraint in Eq.\ref{eq:constraint_part_a}. 

\begin{equation}
 (\mathbf{C2}):  \sum_{\mathbf{h} \in \mathcal{H}_{\mathbf{a}_i}} \mathbf{h}  \le 1 - \mathbf{c}_{\mathbf{a}_i}
\label{eq:constraint_part_a}
\end{equation}
where $\mathcal{H}_{\mathbf{a}_i}$ is the set of all binary heuristic variables for heuristics which use part $\mathcal{A}_i$ and $\mathbf{c}_{\mathbf{a}_i}$ is a binary variable which is equal to $1$ if $\mathcal{A}_i$ has been used in the carbon footprint computation.
Intuitively, if $\mathbf{c}_{\mathbf{a}_i}$ is equal to $1$ then the right-hand side of the constraint is equal to $0$ and no heuristic for $\mathcal{A}_i$ can be selected.
If the carbon footprint of $\mathcal{A}_i$ has not been used for comparison, then $\mathbf{c}_{\mathbf{a}_i}$ is equal to $0$, the right-hand side is equal to $1$ and at most one heuristic for $\mathcal{A}_i$ can be selected.
The effect of this constraint can be observed in \textcolor{blue}{Figure~\ref{fig:example_bip}}, where $\mathbf{h}_{44}$ could not have been selected because $\mathbf{c}_{a_4}$ has already been set to $1$.

Finally, we have to include the carbon footprint comparison.
For each part with complete information, we assign a weight $w_i$ which is its actual carbon footprint.
To show that the carbon footprint of $\mathcal{A}$ is bigger than the carbon footprint of $\mathcal{B}$, we must ensure that the weighted sum of the parts selected for carbon footprint comparison of the former is bigger than the weighted sum of the latter, which is expressed via the constraint in Eq.\ref{eq:constraint_carbon_footprint}.

\begin{equation}
  (\mathbf{C3}):   \sum w_{\mathbf{a}_i} \mathbf{c}_{\mathbf{a}_i} \ge \sum w_{\mathbf{b}_i} \mathbf{c}_{\mathbf{b}_i}
\label{eq:constraint_carbon_footprint}
\end{equation}
In the toy example in Fig.\ref{fig:example_bip}, the left-hand side of this constraint sums up to $20$, which is greater than the right-hand side which selects only $w_{b_3}=10$.

\paragraph{Summary} The objective of the integer program is to maximize the sum of $\mathbf{b}_i$ (Eq.\ref{eq:obj_function}).
However, to assign $\mathbf{b}_i$ to $1$, either a corresponding heuristic variable $\mathbf{h}$ or its carbon footprint variable $\mathbf{c}_{\mathbf{b}_i}$ have to be assigned to $1$ (Eq.\ref{eq:obj_function}).
Whereas a heuristic can only be selected if the counterpart in $\mathcal{A}$ has not already been used by another heuristic (Eq.\ref{eq:constraint_part_a}), the carbon footprint can only be used if we have enough carbon footprint in $\mathcal{A}$ to counterbalance the use of $\mathcal{B}_i$ (Eq.\ref{eq:constraint_carbon_footprint}).
And we ensure that resources in $\mathcal{A}$ are only used once via Eq.\ref{eq:constraint_part_b}.

Note that theoretically, a part $\mathcal{B}_i$ could be shown to be subsumed by $\mathcal{A}$ via multiple heuristic comparisons and/or carbon footprint comparison at the same time, i.e. we do not enforce comparison uniqueness as for $\mathcal{A}_i$ in Eq.\ref{eq:constraint_part_a}.
However, it is not advantageous for the solver to put both kinds of variables for $\mathcal{B}_i$ to $1$, since both would have to be weighed up against by parts in $\mathcal{A}$.

\subsection{Implementation}\label{sec:integer_program_impl}

\begin{figure}
    \centering
    \includegraphics[width=0.8\textwidth]{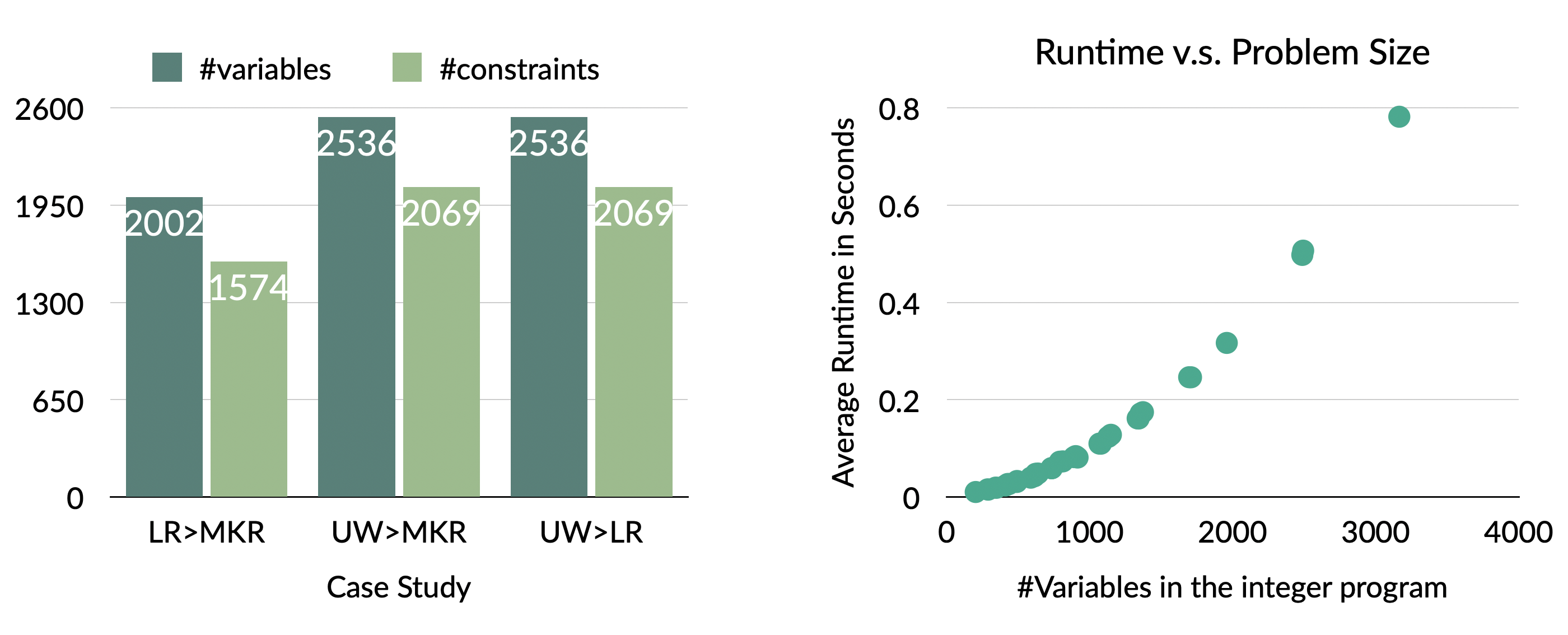}
    \caption{Left: The size of the integer program illustrated by the number of variables and the number of constraints for each of the comparisons in the case study (Section~\ref{sec:complexdesign}) of comparing Arduino dev boards. Right: average runtime in seconds (on a MacBook Pro with M1 chip and 8GB ram) over the size of the problem (represented by number of variables in the integer program).}
    \label{fig:runtime}
\end{figure}

\paragraph{Integer Program}
We implemented the integer program using the Python API of OR-tools\cite{or_tools} and for common comparisons, the runtime is a fraction of a second. Although integer programming is NP-Complete and combinatorial in nature, in practice, solutions to a problem of a reasonable size can be found efficiently. We evaluated the runtime of our comparison algorithm by making pairwise comparisons among eight PCB boards (the three complex designs from the case study (Section~\ref{sec:complexdesign}) and five smaller designs). 
As shown in \textcolor{blue}{Figure~\ref{fig:runtime}}, our comparison algorithm's average runtime for a problem of a size similar to the complex design comparisons (i.e., $\sim 2,500$ variables) is around 0.5 seconds and thus enables our UI to run at interactive speeds.

\begin{figure}[b]
  \centering
  \includegraphics[width=0.8\linewidth]{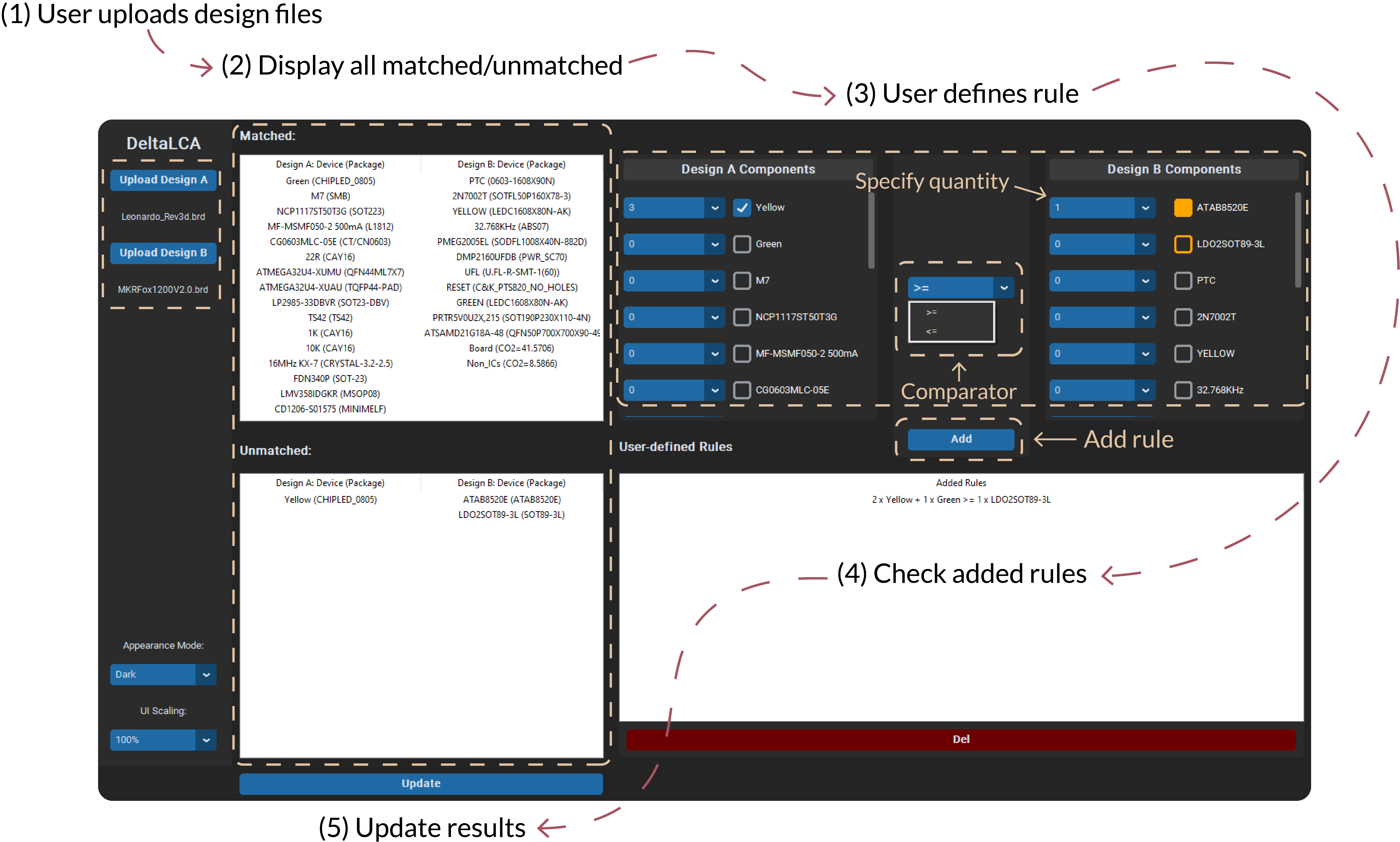}
  \caption{System Screenshot of DeltaLCA user interface. (1) The user uploads design files using the buttons in the sidebar. (2) The user is presented with matched/unmatched components. (3) The user specifies custom comparison rules. (4) The user checks all added rules. (5) Updates the comparison results with user-defined rules taken into account.
  }
  \Description{}
  \label{fig:UI}
\end{figure}

\paragraph{User Interface for Comparing Designs}
The UI for DeltaLCA is developed with Python's tkinter framework (shown in \textcolor{blue}{Figure~\ref{fig:UI}}) and bridges the automated LCI with the comparison algorithm. After the user uploads two design files using the buttons in the sidebar, the system initiates an inventory generation process; the inventory may be partially completed due to the incomplete component information from the design file. Then the initial comparison runs automatically and the user is presented with two tables: one displaying the matched components from designs A and B respectively, and another below listing any unmatched components. The user can analyze individual components in each design via a scrollable frame on the top right. All unmatched components are highlighted in yellow and located at the top of the list to emphasize their selection priority, in contrast to matched components shown in gray. The user can create custom comparison rules by selecting components and specifying their quantities via checkboxes and dropdown menus before the component names, and setting the comparison direction through the comparator option menu. Custom user-defined rules are added to a separate table using the `Add' button, while the `Del' button allows for the removal of specific rules. The user can then click the `Update' button to rerun the comparison algorithm taking into account their custom rules. The comparison process concludes once all components, from the design deemed to have a lower EI, are successfully matched.

\section{Evaluation}
Having demonstrated the technical efficacy of our DeltaLCA automated inventory and comparison algorithm and integrating DeltaLCA with a frontend user interface, we evaluated the performance of our end-to-end system with users in the loop. Our study aimed to answer the following questions:

\begin{itemize}
\item RQ1: Can DeltaLCA accurately calculate the environmental impact (EI) of simple PCB designs with complete information without extra intervention from users?
\item RQ2: To what extent can DeltaLCA compare and analyze complex PCB designs with incomplete component information, and what is the role of the user in this process?
\item RQ3: How do electronics designers perceive the importance of EI in PCB design and what are their attitudes towards incorporating tools like DeltaLCA in their workflow?
\end{itemize}

Below, we detail the methodology and findings of our two-pronged evaluation approach, encompassing both quantitative analysis and qualitative user feedback.

\subsection{Case Study}
Our case study focused on two key scenarios: simple electronic devices, e.g., those with fewer than 50 components, to assess the tool's direct EI calculation capabilities, and more complex PCBs with missing component information to evaluate the tool's comparison efficiency and the degree of user intervention required.

\begin{figure}[b]
  \centering
  \includegraphics[width=\linewidth]{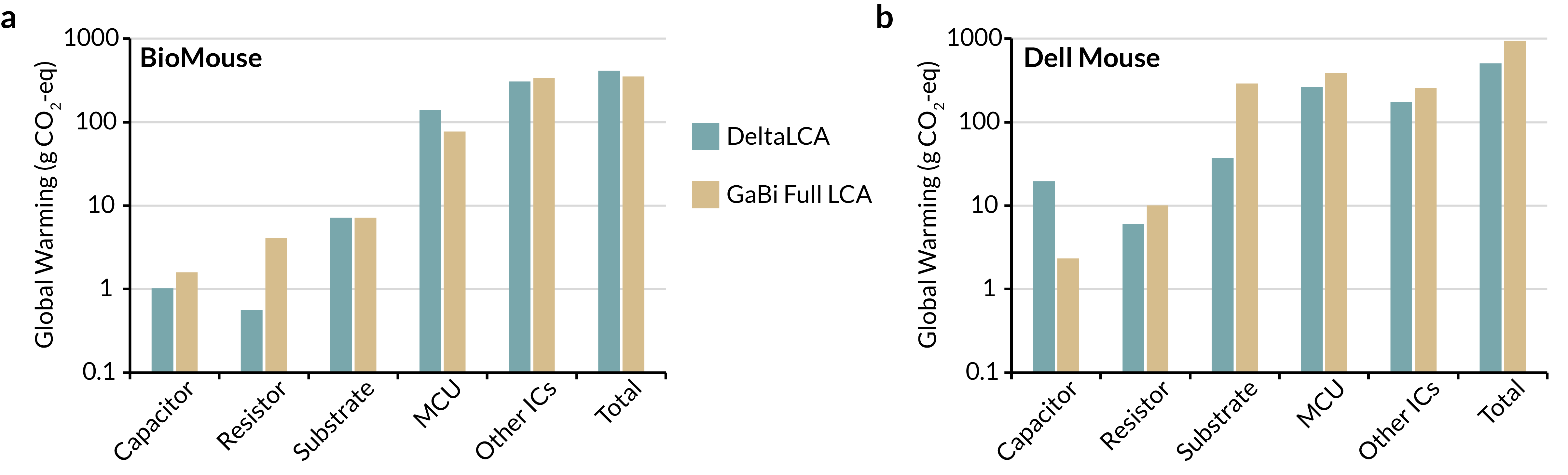}
  \caption{Comparing global warming emissions in carbon dioxide equivalent between DetlaLCA and traditional full LCA using GaBi for a BioMouse (a) and a Dell optical mouse (b).
  }
  \Description{}
  \label{fig:case_study_fullLCA}
\end{figure}

\subsubsection{Complete Inventory with Full LCA (RQ1)}
We employed DeltaLCA to analyze the EI of two real-world designs obtained from the authors of ~\cite{arroyos_tale_2022}: a custom biodegradable mouse and a Dell optical mouse. To benchmark the accuracy of DeltaLCA's results, we compared them with the emissions quantified through a traditional Full LCA using GaBi. DeltaLCA successfully computed the total global warming emissions in carbon dioxide equivalent (CO2-eq) for each design, and determined that BioMouse possesses a lower EI than the Dell mouse (results are shown in \textcolor{blue}{Figure~\ref{fig:case_study_fullLCA}}). The carbon emissions estimated by DeltaLCA are in accordance with GaBi full LCA: the discrepancies stood only at 17.38\% for the BioMouse and -46.73\% for the Dell mouse. In the context of LCA research, such variances are considered within acceptable bounds. We investigate the difference and find that this results from different environmental impact coefficients for components such as PCBs and capacitors between our analyses. We note that our coefficients were determined from publicly available research papers while the LCAs from \cite{arroyos_tale_2022} were from the Gabi Electronics extension database.

Prior studies, such as those by Smith et al. \cite{smith_life_2018} \& Zhang et al. \cite{zhang_comparative_2022} and Liu et al. \cite{liu_future_2014} \& Ozkan et al. \cite{ozkan_life_2018} have acknowledged that LCA results can exhibit up to a three-fold difference even for same products because LCA is susceptible to inconsistencies in methodologies between two studies as we described in Section~\ref{sec:clca} in \textit{Related Work}.

The expert user who performed the LCA from Arroyos et al.~\cite{arroyos_tale_2022} also noted that modeling the the Dell Mouse with standard commercial processes required 2-3 hours while the Biodegradable mouse required 5 hours due to the complexity of modeling a new process flow for the custom water soluble material. This duration only accounts for the time to create the GaBi model and excludes the time required for new component inventory generation. In stark contrast, DeltaLCA achieves a correct comparison result and accelerates the process to a single click.

\subsubsection{Complex Design with Partial Inventory (RQ2)}\label{sec:complexdesign}
In the realm of electronics design, the assessment of more complex devices is crucial, as they typically incorporate a diverse range of components, and have incomplete component information occurring in their design. To illustrate the benefit of our comparative assessment tool, we did a case study for comparing the carbon footprint of three Arduino dev boards: (1) Arduino Leonardo R3, (2) Arduino MKR FOX 1200, and (3) Arduino UNO Wifi. We choose these devices forseveral reasons: 1) Complexity: Unlike simpler devices like the mouse above, these development boards are akin to small computers, featuring intricate circuitry and multiple-layer design. 2) Relevance in UbiComp: these platforms are widely used for prototyping and educational purposes.

Our study involved a cross-comparison of these three devices. One expert with five years of electronics design experience, but no LCA experience, used our tool to establish a relative EI ranking of the three boards, namely $(3) \geq (1) \geq (2)$. For reference, we also collected some basic statistics about the three boards in \textcolor{blue}{Table A\ref{tab:boards}}. If we only deduce the relative ordering based on the information in this table, it is not obvious what the relative order should be. In fact, the expert user initially tried to prove (2) $\geq$ (1) with our UI but in the end could not figure out a way to do so and realized that it should be in the opposite direction. This demonstrates that our tool could also serve as a verifier to help users verify their speculation.

\begin{figure}
    \centering
    \includegraphics[width=0.9\textwidth]{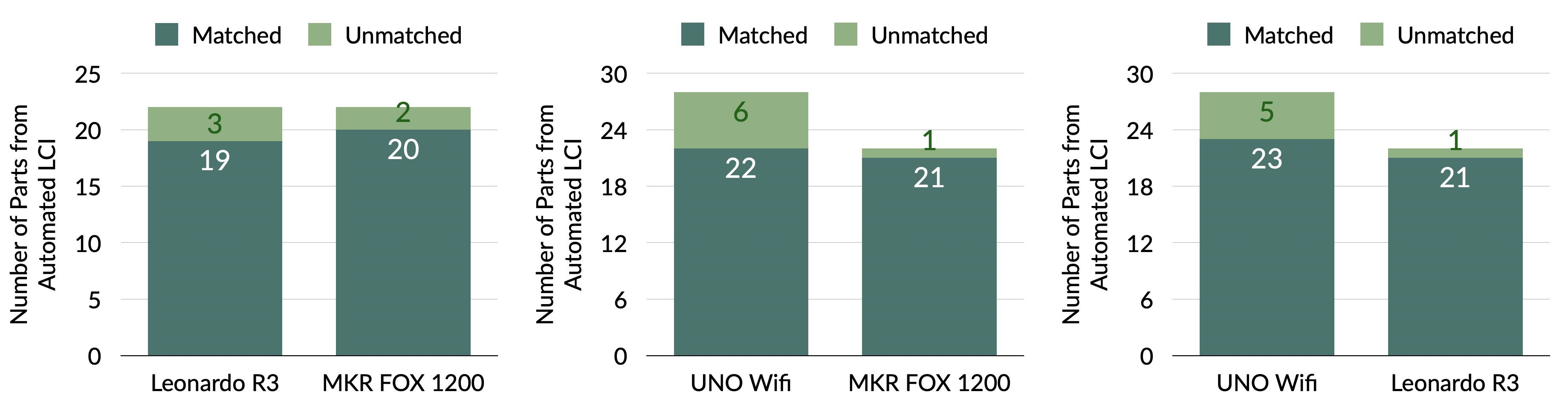}
    \caption{The matching coverage of our comparison algorithm without any user-defined rules for the three comparisons proving that design A's EI on the left is greater than design B's EI on the right. Our method can automatically find a valid matching that covers most of the parts in design A and design B. Here the parts refer to the list generated by the automated life-cycle inventory process where we turn components like resistors, capacitors, and inductors into a single component `NonICs' and the base board itself into a single component `Board'.}
    \label{fig:matchingcoverage}
\end{figure}

As shown in \textcolor{blue}{Figure~\ref{fig:matchingcoverage}}, for complex devices like these, containing around 100 components and 20$\sim$30 ICs, our tool's comparative algorithm can automatically match most components (on average 88\% overall, and 94\% for the target, design B in this case) without any user intervention. 

\textcolor{blue}{Figure A~\ref{fig:casestudy3}} shows that typically fewer than five user-defined rules (specifically, 4, 2, and 1 for the three comparisons) are needed to complete a comparison, with an average time spent of less than 15 minutes per comparison. This is a significant reduction in time compared to traditional LCAs for simpler designs (mouse examples in the previous section). Additionally, our tool does not require LCA specific expertise, empowering electronics designers to perform evaluations directly. Such efficiency demonstrates the substantial time savings DeltaLCA offers designers.


In our study on complex designs with partial inventories, we highlight the strengths of our comparison algorithm which can utilize both carbon footprint data from automated LCI when available, and also apply expert-defined heuristics when only partial information exists. Moreover, users have the flexibility to create rules for any components across the designs, not just the unmatched ones. This feature is crucial because the matching derived from the integer program may not always be globally optimal and users might have insights on specific components that aren't automatically extracted from PCB design files and online data sources.

\begin{figure}[b]
  \centering
  \includegraphics[width=\linewidth]{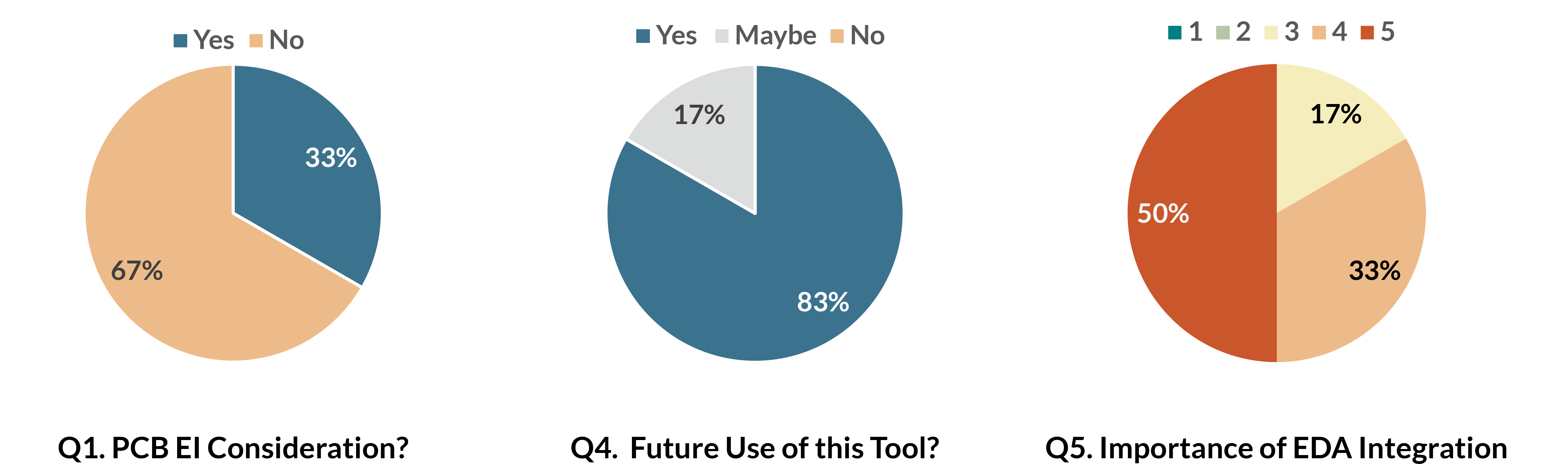}
  \caption{Left: Distribution of users who have considered the environmental impact when designing PCB. Middle: Distribution of users who are inclined to incorporate DeltaLCA into the future design workflow. Right: Distribution of users who think it is important to plug DeltaLCA into existing EDA software, on a scale of 1 to 5 where 1 represents the least importance and 5 represents the most importance.
  }
  \Description{}
  \label{fig:survey_result}
\end{figure}

\subsection{User Evaluation \& Survey (RQ3)}
To deepen our understanding of the tool’s utility and effectiveness in real-world scenarios, we conducted user evaluations to understand how designers percieve sustainable design and our tool. The study is composed of two parts. First, we provided a brief background introduction about EI of electronics and guided the participants through a demonstration of DeltaLCA. Second, participants filled out a survey. Please refer to \textcolor{blue}{Table A\ref{tab:survey}} for the complete list of our survey questions. It is noteworthy that this study aims to explore designers' awareness of environmental considerations in their workflows rather than performing comprehensive usability testing of  DeltaLCA. Furthermore, we sought to gauge the likelihood of DeltaLCA's integration into designers' workflows.


For our study, we recruited 6 participants: 2 aged 18-24 and 4 aged 25-34; 5 research assistants at university and 1 hardware architect in industry; 2 women and 4 men. We present the summarized outcomes of our user survey below (\textcolor{blue}{Figure \ref{fig:survey_result}}).

\subsubsection{PCB Environmental Impact consideration}
The first aspect of our survey investigated whether PCB designers have previously considered EI during their design process. Our results indicate a relatively low level of environmental consideration: only 33.33\% reported having considered EI when designing PCBs, while the majority 66.67\% had not. One of the participants who has considered EI mentioned: \textit{``but more on the side of rare metals, not carbon footprint.''}

\subsubsection{Future use of this tool}
The survey results revealed a strong inclination among participants towards incorporating DeltaLCA into their future design processes, while only one participant indicated uncertainty (`Maybe'). Significantly, there was no outright rejection of DeltaLCA, indicating a general openness to its adoption within the design community.

\textbf{Importance of EDA integration.} 
Even so, we noticed how crucial it is for DeltaLCA to plug in existing EDA software, as this could impact its usability and convenience. When asked about the importance of DeltaLCA’s integration with existing EDAs, the majority of respondents (83.33\%) rated this aspect as highly crucial (scores of 4 and 5). This finding aligns with one participant's note, \textit{``going out of the way to evaluate this kind of an issue is too hard, but if it's built-in, I can easily check with one button.''}			

\textbf{Desired features.} 
When asked about additional desired functionalities in a tool designed for assessing the environmental impact of PCBs, more than half of the participants mentioned \textit{``automatic [design/component] optimization [strategies/guidelines/suggestions] for reducing environmental impact.''}
\section{Discussion \& Conclusion}

In this paper, we introduce an innovative method for creating design tools that facilitate environmentally-focused decision-making. This approach is grounded in the key understanding that, although Life Cycle Assessment (LCA) calculations are complex, comparative analysis can enable effective decision-making even with incomplete data. Our end-to-end tool, specifically developed for electronics design, demonstrates that leveraging domain-specific knowledge and datasets can significantly automate the process and require minimal user input. The tool's effectiveness was validated through case studies and a user survey.

While our work introduces a new direction in this field, it also uncovers multiple opportunities for future research.First, there is considerable potential to enhance the system. Improvements could be made to the Automated LCI, especially by employing novel search algorithms, with advancements in Large Language Models (LLMs) offering possibilities to explore broader databases or to identify freely available web data.  In regards to the matching algorithm, although our present methodology effectively combines complete and partial data, thereby reducing unmatched components, a re-evaluation of our objectives could be more beneficial. The primary goal should shift towards reducing the number of additional user rules required for making final decisions, rather than just minimizing the count of unmatched components. Currently, the number of matches is used as a proxy for this goal, but future work may reveal more effective methods, possibly through deeper domain expertise or advanced data analysis.

Secondly, there is a wealth of opportunities to further develop our insights on comparative approaches to LCA. Future research could adapt this methodology to other domains, such as architecture, material design, and garment production, among others. Additionally, it would be intriguing to explore alternative systems that incorporate comparative assessments at different stages of the design pipeline. Although our current research focuses on comparing two final designs, the concept of early-stage comparisons in the design process is a promising area for future work. Such endeavors would likely require not only heuristic but also probabilistic reasoning.

In summary, this work introduces a novel approach to sustainable design with multiple avenues for further exploration. We hope that our research will both immediately enable designers in the Ubicomp community to develop more sustainable design practices and encourage further investigation into LCA-focused design tools.

\bibliographystyle{ACM-Reference-Format}
\bibliography{references}

\appendix
\clearpage
\section{Case Study Summary}
We summarized the basic attributes of the three Arduino dev boards and user interaction with our tool for comparing the EIs of these devices in the case study.

\begin{table}
    \centering
    \begin{tabular}{c|c|c|c|c|c}
    & \textbf{Board} & \textbf{Board Area (mm${}^2$)} & \textbf{\#Layers} & \textbf{\#Components} & \textbf{\#ICs from Automated LCI}\\
    \hline
    (1) & Leonardo R3 & 4180.6368 & 2 & 97 & 20\\
    (2) & MKR FOX 1200 & 1696.76 & 4 & 92 & 20 \\
    (3) & UNO Wifi & 3659.124 & 2 & 117 & 26
    \end{tabular}
    \caption{Basic attributes of the three Arduino dev boards.}
    \label{tab:boards}
\end{table}

\clearpage
\begin{figure}
     \centering
     \begin{subfigure}[b]{0.85\textwidth}
         \centering
        \includegraphics[width=\textwidth]{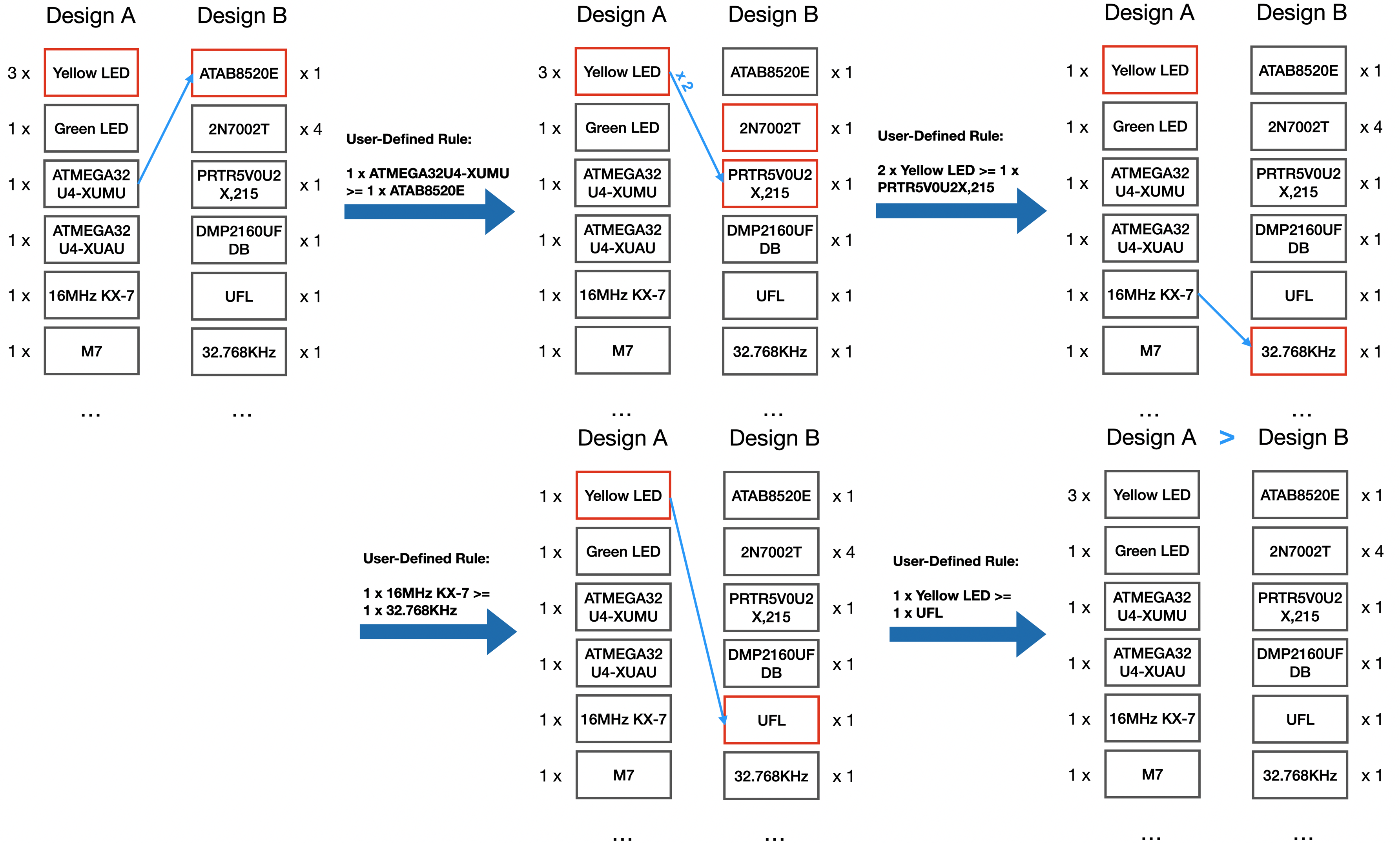}
         \caption{Arduino Leonardo R3 $\geq$ Arduino MKR FOX 1200}
         \label{fig:MKRgtLR}
     \end{subfigure}
     \\
     \begin{subfigure}[b]{0.85\textwidth}
         \centering
         \includegraphics[width=\textwidth]{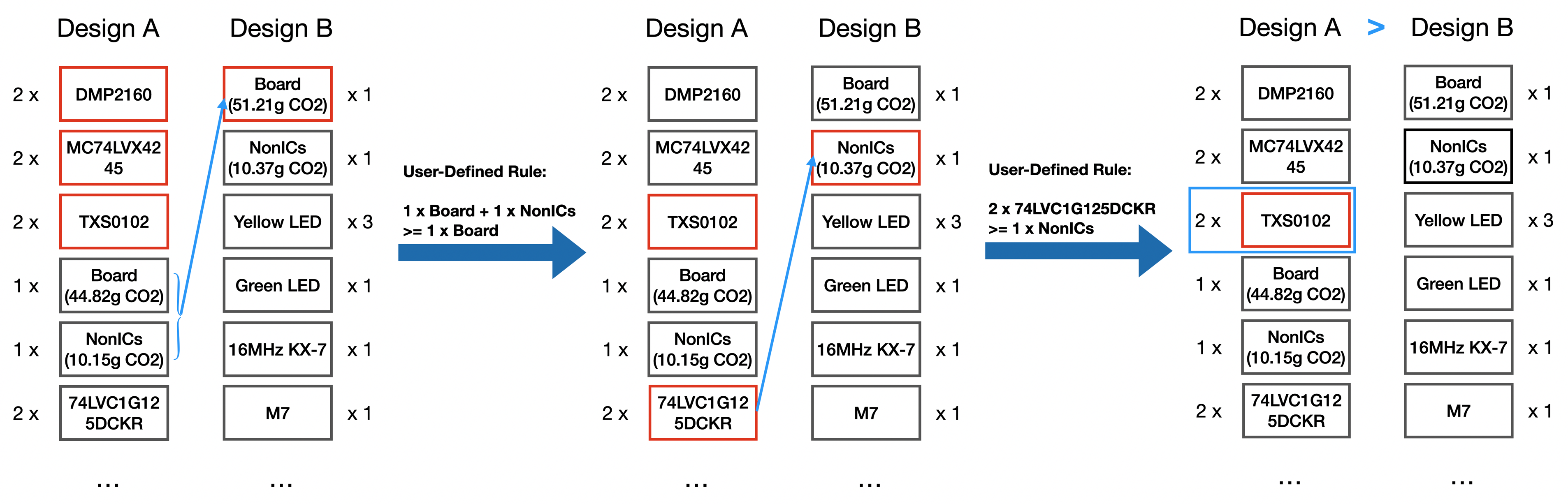}
         \caption{Arduino UNO Wifi $\geq$ Arduino MKR FOX 1200}
         \label{fig:UWgtMKR}
     \end{subfigure}
     \\
     \begin{subfigure}[b]{0.51\textwidth}
         \centering
         \includegraphics[width=\textwidth]{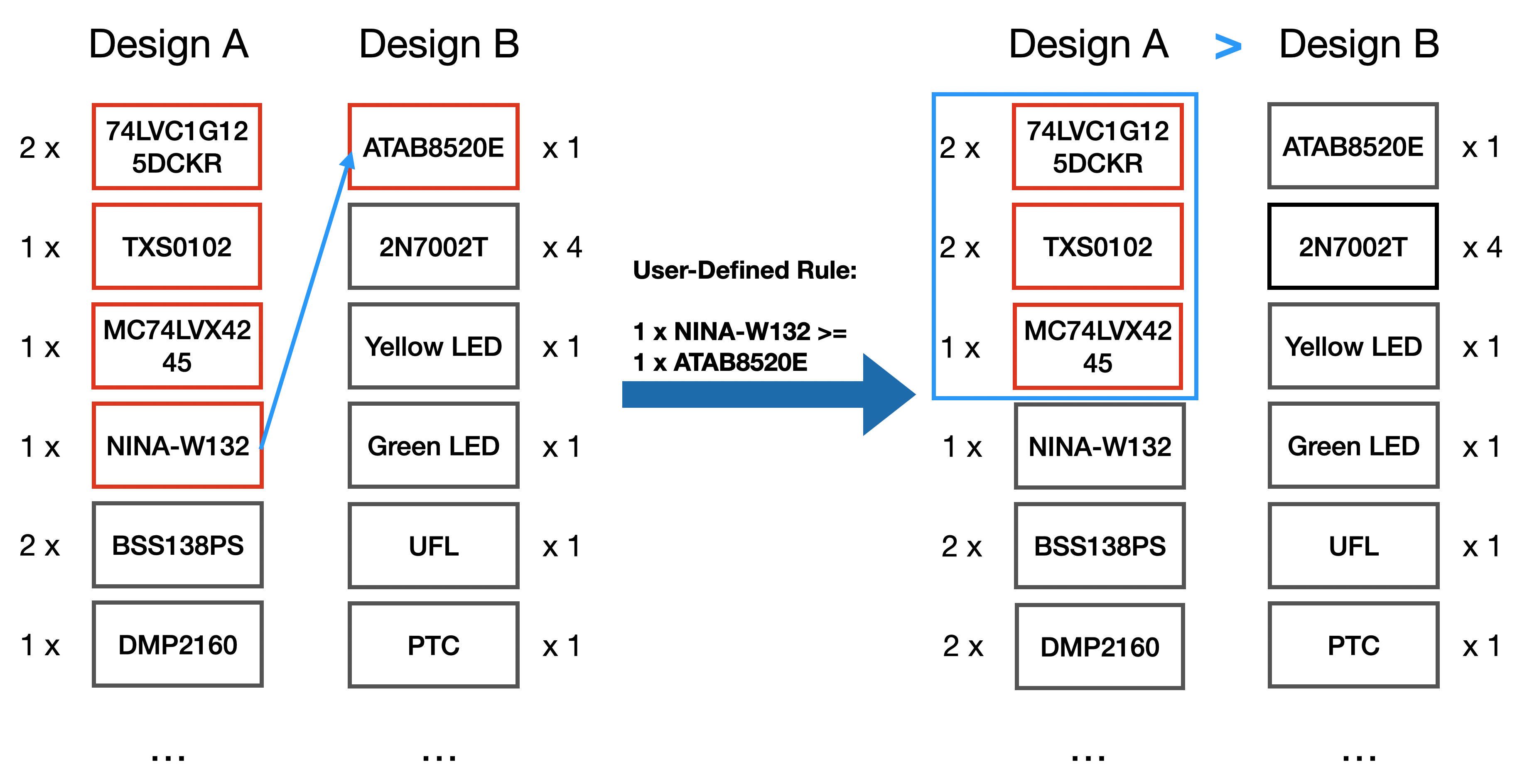}
         \caption{Arduino UNO Wifi $\geq$ Arduino Leonardo R3}
         \label{fig:UWgtLR}
     \end{subfigure}
        \caption{Summary of the user interactions with the UI for comparing the three boards. In the subfigures, the red boxes represent unmatched components, and the gray boxes represent matched components. The blue arrow indicates a user-defined rule between components in design A and design B. Here the user is proving A > B for all the three cases, so they stop when all components in B are matched and zero or more components in A are unmatched (indicated with the blue box in the last step).}
        \label{fig:casestudy3}
\end{figure}

\clearpage

\section{User Evaluation \& Survey Results}

\begin{table}[h]
\begin{tabular}{@{}rp{12cm}@{}}
\toprule
\multicolumn{1}{l}{\#} &
  Question \\ \midrule
\rowcolor[HTML]{FFFFFF} 
1 &
  Have you ever considered the environmental impact when designing a Printed Circuit Board (PCB)? \\
  
\rowcolor[HTML]{F3F3F3} 
2 &
  In your opinion, what strategies or approaches would you employ to design a PCB to minimize its carbon footprint? \\

\rowcolor[HTML]{FFFFFF} 
3 &
  Please indicate your perception of the factors that significantly influence the carbon footprint of a PCB by ranking the following factors from most important to least important: a. Die area of the Integrated Circuit (IC). b. Process technology node of IC. c. Package size of IC (overall size). d. Board size. e. Number of components. f. Number of PCB layers. \\

\rowcolor[HTML]{F3F3F3} 
4 &
  Would you consider incorporating this tool into your PCB design workflow? \\

\rowcolor[HTML]{FFFFFF} 
5 &
  On a scale of 1 to 5, where 1 represents the least importance and 5 represents the most importance, how crucial is it for this tool to integrate with existing Electronic Design Automation (EDA) software (like EAGLE or KiCAD)?  \\
  
\rowcolor[HTML]{F3F3F3} 
6 &
 What additional features or functionalities would you desire in a tool designed for assessing the environmental impact of PCBs? \\ \bottomrule
\end{tabular}
\caption{List of the survey questions.}
\label{tab:survey}
\end{table}

\end{document}